\documentclass[sigconf]{acmart}

\usepackage{epsfig,endnotes}
\usepackage[utf8]{inputenc}

\usepackage{amsmath}
\usepackage{amsfonts}
\usepackage{amssymb}
\usepackage{amsthm}

\usepackage{bbm}
\usepackage{mathtools}
\usepackage[nounderscore]{syntax}
\usepackage{algorithm}
\usepackage{algpseudocode}
\usepackage{graphicx}

\usepackage{subcaption}
\usepackage{stmaryrd}
\usepackage{url}

\usepackage{standalone}
\usepackage{cancel}
\usepackage{listings}
\usepackage{xspace}

\usepackage{tikz}
\usetikzlibrary{
  shapes,calc,arrows,fit,positioning,decorations.pathmorphing,snakes,intersections,shapes.geometric,trees
}

\newcommand{\cut}[1]{ }

\newcommand{\assocmeasure}[0]{\ensuremath{d}}
\newcommand{\infmeasure}  [0]{\ensuremath{\iota}}

\newcommand{\utilmeasure} [0]{\ensuremath{v}}

\newcommand{\proxyusage}   [0]{Proxy Use\xspace}

\newcommand{\argproxyusage}[2]{$(#1,#2)$-\proxyusage}
\newcommand{\edproxyusage} [0]{\argproxyusage{\epsilon}{\delta}}

\newcommand{\proxyuse}   [0]{proxy use\xspace}

\newcommand{\argproxyuse}[2]{$(#1,#2)$-\proxyuse}
\newcommand{\edproxyuse} [0]{\argproxyuse{\epsilon}{\delta}}
\newcommand{\Proxyuse}   [0]{Proxy use\xspace}

\newcommand{\ProxyUse}   [0]{Proxy Use\xspace}

\newcommand{\indicator}[1]{\mathbbm{1}\paren{#1}}

\newtheorem{thm}   {Theorem}
\newtheorem{defn}  {Definition}
\newtheorem{property}{Property}

\makeatletter
\def\thm@space@setup{\thm@preskip=1pt
\thm@postskip=1pt}
\makeatother

\newcommand{\code}[1]{\lstinline!#1!}

\newcommand{\stacklabel}[1]{\stackrel{\smash{\scriptscriptstyle \mathrm{#1}}}}
\newcommand{\defeq}{\stacklabel{def}=}

\newcommand{\sref}[1]{\xspace{}Section \ref{#1}}

\newcommand{\paren}[1]{\left( #1 \right)}
\newcommand{\sparen}[1]{\left[ #1 \right]}
\newcommand{\set}[1]{\left\{ #1 \right\}}

\DeclareMathOperator*{\E}{\mathbb{E}}
\newcommand{\prob}[1]{\Pr\paren{#1}}

\newcommand{\expectsub}[2]{\E_{#1}\sparen{#2}}

\newcommand{\features}{\mathbf{X}}

\newcommand{\samp}{\ensuremath{\stacklabel{\$}\gets}}

\newcommand{\cal}{\mathcal}
\newcommand{\oracle}{{\cal O}}
\newcommand{\model}{{\cal A}}
\newcommand{\population}{{\cal P}}

\newcommand{\denote}[1]{\llbracket #1 \rrbracket}

\newcommand{\subst}[3]{[#1/#2]#3}
\newcommand{\pair}[2]{\langle #1, #2\rangle}
\newcommand{\pairpop}[2]{\pair{#1}{#2}_{\population}}
\newcommand{\pairpopbrace}[2]{\pair{\{#1\}}{#2}_{\population}}

\DeclareMathOperator*{\argmax}{arg\,max}
\DeclareMathAlphabet{\mathcal}{OMS}{cmsy}{m}{n}  %

\definecolor{mygray}{rgb}{0.5,0.5,0.5}
\lstdefinestyle{customlisp}{
  belowcaptionskip=1\baselineskip,
  breaklines=true,
  language=Lisp,
  showstringspaces=false,
  numbers=left,
  xleftmargin=2em,
  framexleftmargin=1.5em,
  numbersep=5pt,
  numberstyle=\tiny\color{mygray},
  basicstyle=\small\ttfamily,
  keywordstyle=\color{blue},
  commentstyle=\itshape\color{purple!40!black},
  stringstyle=\color{orange},
  morekeywords={def-rule, ?},
  tabsize=2
}

\makeatletter
\renewcommand{\paragraph}{%
  \@startsection{paragraph}{4}%
  {\z@}{0.75ex \@plus 0.5ex \@minus .2ex}{-1em}%
  {\normalfont\normalsize\bfseries}%
}
\makeatother

\usepackage{enumitem}
\setlist{nosep}

\algtext*{EndFor}
\algtext*{EndIf}
\algtext*{EndProcedure}

\title{{Proxy Discrimination\footnotemark{} in Data-Driven
    Systems}\\{\Large Theory and Experiments with Machine Learnt
    Programs}}

\author{
{\rm Anupam Datta} \\
CMU
\and
{\rm Matt Fredrikson} \\
CMU
\and
{\rm Gihyuk Ko} \\
CMU
\and
{\rm Piotr Mardziel} \\
CMU
\and
{\rm Shayak Sen} \\
CMU
}

\begin{abstract}
  Machine learnt systems inherit biases against protected classes,
  historically disparaged groups, from training data. 
  Usually, these biases are not explicit, they rely on subtle correlations
  discovered by training algorithms, and are therefore difficult to
  detect.
  We formalize a notion of \emph{proxy discrimination} in data-driven systems, a
  class of properties indicative of bias, as the presence of protected
  class correlates that have causal influence on the system's output.
  We evaluate an implementation on a corpus of social datasets,
  demonstrating how to validate systems against these properties and
  to repair violations where they occur.
\end{abstract}


\begin{document}

\maketitle

\footnotetext[1]{This work is a companion paper to an earlier
  paper\cite{useprivacy} where the same techniques presented here were
  applied to formalizing and enforcing privacy restrictions.}

\section{Introduction}
\label{sect:intro}

Utility of machine learning has spurred adoption of automated systems
in many areas of life. 
Systems, from credit and insurance assessors\cite{marr15forbes} to
recidivism predictors\cite{angwin16propublica}, have significant
impact on the affected individuals' future. 
Machine learnt systems, however, are constructed on the basis of
observational data from the real world, with its many historical or
institutionalized biases. 
As a result, they inherit biases and discriminatory practices inherent
in the data. 
Adoption of such systems leads to unfair outcomes and the perpetuation
of biases.

Examples are plentiful: race being associated with predictions of
recidivism~\cite{angwin16propublica}; gender affecting displayed
job-related ads~\cite{datta15pets}; race affecting displayed search
ads~\cite{sweeney13cacm}; Boston's Street Bump app focusing pothole
repair on affluent neighborhoods~\cite{rampton14reuters}; Amazon's
same day delivery being unavailable in black
neighborhoods~\cite{ingold16bloomberg}; and Facebook showing either
``white'' or ``black'' movie trailers based upon ``ethnic
affiliation''~\cite{statt16verge}.

Various instances of discrimination are prohibited by law.
In the United States, for example, Title VII of U.S. 
Civil Rights act prohibits making employment decisions on the basis of
race, sex, and other protected attributes~\cite{title-vii}. 
Further legislation makes similar restrictions on the use of protected
attributes for credit~\cite{fed06book} and housing
decisions~\cite{kennedy15scotus}. 
Other law establish similar protections in other jurisdictions
\cite{ukequalityact2010}.

In the United States, legal arguments around discrimination follow one
of two frameworks: \emph{disparate treatment} or \emph{disparate
  impact}~\cite{barocas2016big}.
Disparate treatment is the intentional and direct use of a protected
class for a prohibited purpose.
An example of this type of discrimination was argued in McDonnell
Douglas Corp.
v.
Green~\cite{mcdonnell}, in which the U.S. 
Supreme Court found that an employer fired an employee on the basis of
their race.
An element of disparate treatment arguments is an
establishment of the protected attribute as a \emph{cause} of the
biased decision~\cite{datta2015plsc}.

Discrimination does not have to involve a direct use of a protected
class; class memberships may not even take part in the decision.
Discrimination can also occur due to correlations between the
protected class and other attributes.
The legal framework of \emph{disparate impact}~\cite{disparate-impact}
addresses such cases by first requiring significantly different
outcomes for the protected class, regardless of how the outcomes came
to be.
An association between loan decisions and race due to the use of
applicant address, which itself is associated with race, is an
example~\cite{hunt2005} of this type of discrimination, as are most of
the examples cited earlier in this introduction.
The association requirement is not causal and thus further arguments
must be made to establish that the cause of the observed disparate
impact can be attributed to use of an attribute correlated with the
protected class, which cannot be excused due to business necessities.

Discrimination arising due to use of features correlated to protected
classes is referred to as \emph{discrimination by proxy} in U.S. 
legal literature or \emph{indirect discrimination} in other
jurisdictions such as the U.K. 
\cite{ukequalityact2010}.
In this paper we will use the term ``proxy'' to refer to a feature
correlated with a protected class whose use in a decision procedure
can result in indirect discrimination \footnote{ This convention
  differs slightly from its use in statistics~\cite{statistics} and
  some legal literature in the U.S. 
  which emphasize that a proxy stands in place for some unobservable
  (or difficult to observe) feature. 
  This corresponds to early U.S. 
  case-law on discrimination: in E.G. 
  Griggs v. 
  Duke Power Co.~\cite{disparate-impact}, for example, a company was
  found to discriminate against black employees by requiring a
  high-school diploma as an indicator of future competency in high
  positions, when in reality high-school diploma had little to do with
  competency in those particular positions.
  In our work we do not demand that proxies are used because some
  target variable is difficult to observe.
}. 
This terminology is consistent with its use in recent works in the 
field of fairness in machine learning 
\cite{fairtest, adler2016auditing, kilbertus2017avoiding}.

In the context of machine learnt systems, non-human decision makers,
direct discrimination as in disparate treatment is not difficult to
recognize: a feature indicating the protected class is present and
used.
Usage can be determined by inspecting the system code or determined
experimentally~\cite{qii}.
Indirect discrimination like disparate impact can also be computed
experimentally.
Such a check, however, does not establish cause which 1) is an element of
legal arguments, 2) underlies business necessity claims which excuse
certain types of disparate outcomes, and 3) is suggestive of remedies
for the repair of bias.

In this work we formalize a notion of proxy discrimination that
captures use of proxies of protected information types (e.g., race,
gender) in data-driven systems. 
Further we design, implement, and apply algorithms for detecting these
types of indirect discrimination and for removing them from machine
learnt models.

\paragraph*{\Proxyuse}
A key technical contribution in this paper is a formalization of
\emph{proxy use} of features in programs, the formal models for
machine learnt systems.
The formalization relates proxy use to intermediate computations
obtained by decomposing a program. 
We begin with a qualitative definition that identifies two essential
properties of the intermediate computation (the proxy): \textit{1)}
its result perfectly predicts the protected information type in
question, and \textit{2)} it has a causal affect on the final output
of the program.

We arrive at this program-based definition after a careful examination
of the space of possible definitions. 
In particular, we prove that it is impossible for a purely semantic
notion of intermediate computations to support a meaningful notion of
proxy use as characterized by a set of natural properties or axioms
(Theorem~\ref{thm:sem-impossibility}). 
The program-based definition arises naturally from this exploration by
replacing semantic decomposition with decompositions of the program. 
An important benefit of this choice of restricting the search for
intermediate computations to those that appear in the text of the
program is that it supports natural algorithms for detection and
repair of proxy use. 
Our framework is parametric in the choice of a programming language in
which the programs (e.g., machine learnt models) are expressed and the
population to which it is applied. 
The choice of the language reflects the level of white-box access that
the analyst has into the program.

Every instance of proxy use does not constitute a case for
discrimination by proxy, and the distinction is a normative judgement.
For example, in the U.S., voluntary attributes such as hair style are
not considered proxies even when they are highly correlated with
race~\cite{rich2004performing}. 
Further, business necessities may excuse the use of even involuntary
(race, gender, etc.) 
proxies.
As a result, our theory relies on a normative judgement oracle to
decide whether a particular proxy use is acceptable. 
\sref{sec:proxy-discrimination} discusses the role of normative
judgement in our theory.

\paragraph*{Closely related work}
A wide body of work addresses the problem of finding discrimination in
machine learning systems and avoiding violations with adjustments to
training data, training algorithms, or trained models (see
\sref{sec:related} for a brief overview).
Threats to fairness from proxy use are recognized in the
literature~\cite{title-vii, dwork12itcs,
  kilbertus2017avoiding, adler2016auditing}. Our treatment 
differs significantly from the prior work. 

Tram{\`{e}}r et al.
developed a system for discovering associations, or proxies in
observational data~\cite{fairtest}. 
Their work emphasizes need for differing association metrics and
provides a means of discovering unwarranted associations in
sub-populations of individuals. 
Both right metrics and right sub-populations are necessary for
discovering and tracking down subtle biases. 
These elements are complementary to our goals and methods. 

Adler et al.~\cite{adler2016auditing} describe a method for estimating
the indirect influence of a protected class on a model's outcome by
computing that model's accuracy on a dataset in which proxies of the
protected class have been obscured. 
They argue that the difference between this accuracy and accuracy on
the un-obscured data is a measure of a proxy's influence in a model
and can determine whether it is a cause of disparate outcomes arising
from indirect use of protected classes.
Their technique is designed not to rely on white-box access to the
models but in order to obscure proxies it assumes that the
relationship between potential proxies and the other attributes can be
learned by a chosen set of algorithms.
Our setting and assumptions differ in that we make no assumptions
about the proxy-attributes relationship (and our notion of association
is information theoretic) though we require white-box access.
We also provide repair algorithms that can strip proxy use from
previously learnt models.

Kilbertus et al.~\cite{kilbertus2017avoiding} follow Pearl's work on
the discrimination~\cite{CausalityBook} by describing indirect/proxy
discrimination in terms of causal graphs.
They also discuss algorithms for avoiding such discrimination in some
circumstances. 
Our work does not require access to a causal graph that specifies
causal relationships between features. 

\paragraph*{Contributions}

We make the following contributions:

\begin{itemize}
\item An conception of \emph{proxy discrimination} in data-driven
  systems that restricts the use of protected classes and some of
  their proxies (i.e., strong predictors) in automated decision-making
  systems (\sref{sec:proxy-discrimination}).
\item A formal definition of \emph{proxy use}---the key building block
  for proxy discrimination--and an axiomatic basis for this definition
  (\sref{sec:defn}).
\item An evaluation of the techniques to several use-cases based on
  real-world datasets.
\end{itemize}

This paper is a companion to our earlier work\cite{useprivacy} where
we motivated and applied proxy use for formalizing and enforcing
privacy restrictions.
We replicate here the motivating and definitional aspects of the
earlier work from a discrimination perspective.
We summarize the algorithms for detection and repair of proxy use but
leave off implementation details and proofs justifying their
suitability.
The earlier work includes an evaluation over privacy-related use-cases
while in this paper we conclude with discrimination-based use cases.
In the earlier paper~\cite{useprivacy} we describe:

\begin{itemize}
\item Algorithmic details of the detection procedure and proof that it
  is sound and complete relative to our proxy use definition.
\item Algorithmic details of a repair algorithm and proof that it
  removes violations of the proxy use instantiation in a machine
  learning model that are identified by our detection algorithm and
  deemed inappropriate by a normative judgment oracle.
\item Implementation details and evaluation of our approach on popular
  machine learning algorithms, including decision trees, random
  forests, and logistic regression, applied to privacy use-cases based
  on real-world datasets.
\end{itemize}

\section{Proxy Discrimination}
\label{sec:proxy-discrimination}

We model the data processing system as a program $p$. 
The proxy discrimination constraint governs a protected class $Z$. 
Our definition of proxy discrimination makes use of two building
blocks: (1) a function that given $p$, $Z$, and a population
distribution $\population$ returns a witness $w$ of proxy use of $Z$
in a program $p$ (if it exists); and (2) a normative judgment oracle
$\oracle(w)$ that given a specific witness returns a judgment on
whether the specific proxy use is appropriate ({\sc true}) or not
({\sc false}). 

Not all instances of proxy use of a protected class  are inappropriate. 
For example, business necessity allows otherwise prohibited uses in some
cases. Our theory of proxy discrimination makes use
of a normative judgment oracle that makes this inappropriateness
determination for a given instance.

\begin{defn}[Proxy Discrimination] Given a program $p$, protected
  class $Z$, normative judgment oracle $\oracle$, and population
  distribution $\population$, a program $p$ exhibits proxy
  discrimination if there exists a witness $w$ in $p$ of \proxyuse of
  $Z$ in $\population$ such that $\oracle(w)$ returns {\sc false}.
\end{defn}

In this paper, we formalize the computational component of the above
definition, by formalizing what it means for a model to use a
protected class directly or through proxies (\S\ref{sec:defn}), and
designing algorithms to detect proxy uses in programs and remove
inappropriate uses (\S\ref{sect:detection-and-repair}). 
We assume that the normative judgment oracle is given and use it to
identify inappropriate proxy uses.
In our experiments, we illustrate our analysis and repair
algorithms to identify proxies and repair ones deemed inappropriate by
the oracle (\S\ref{sect:eval}).

The normative oracle separates computational considerations that are
mechanically enforceable and ethical judgments that require input from
human experts. 
This form of separation exists also in some prior work on fairness
~\cite{Dwork12} and privacy~\cite{garg-jia-datta-ccs-2011}.

\section{Proxy Use: A Formal Definition}
\label{sec:defn}

We now present an axiomatically justified, formal definition of proxy
use in data-driven programs.
Our definition for proxy use of a protected class involves
\emph{decomposing} a program to find an intermediate computation whose
result exhibits two properties:
\begin{itemize}
  \item \emph{Proxy}: strong association with the protected class
  \item \emph{Use}: causal influence on the output of the program
\end{itemize}

In \S~\ref{sec:defn:examples}, we present a sequence of examples to
illustrate the challenge in identifying proxy use in systems that
operate on data associated with a protected class.
In doing so, we will also contrast our work with closely-related work
in privacy and fairness.
In \S\ref{sec:notation}, we formalize the notions of proxy and use,
preliminaries to the definition.
The definition itself is presented in \S\ref{sec:definition} and
\S\ref{sec:defn:quant}.
Finally, in \S\ref{sec:defn:impossibility}, we provide an axiomatic
characterization of the notion of proxy use that guides our
definitional choices.
We note that readers keen to get to the discussion of the detection
and repair mechanisms may skip \S\ref{sec:defn:impossibility} without
loss of continuity.

\subsection{Examples of Proxy Use}
\label{sec:defn:examples}

Prior work on detecting use of protected information
types~\cite{datta2015opacity,sunlight,fairtest,feldman15disparate} and
leveraging knowledge of detection to eliminate inappropriate
uses~\cite{feldman15disparate} have treated the system as a black-box.
Detection relied either on experimental access to the
black-box~\cite{datta2015opacity,sunlight} or observational data about
its behavior~\cite{fairtest,feldman15disparate}.
Using a series of examples demonstrating redlining\cite{}, we motivate
the need to peek inside the black-box to detect proxy use.

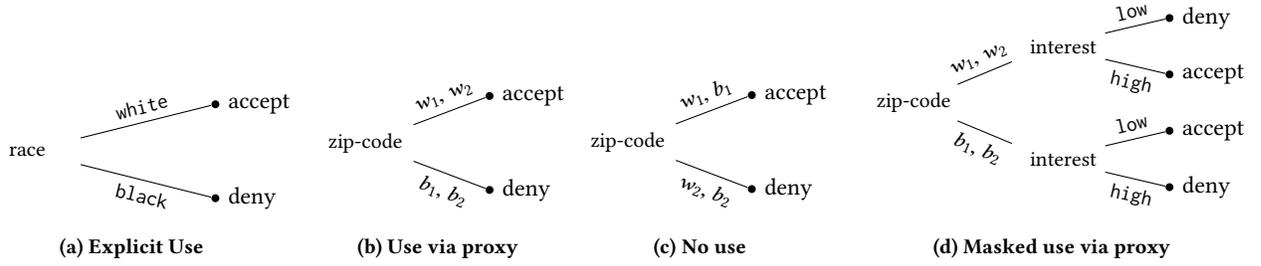
\begin{figure*}[t]
\centering

\tikzstyle{bag}=[text width=4em,text centered,font=\small]
\tikzstyle{end}=[circle,minimum width=3pt,fill,inner sep=0pt]

\tikzstyle{level 1}=[level distance=2.5cm, sibling distance=1.25cm]
\tikzstyle{level 2}=[level distance=2.5cm, sibling distance=0.75cm]
\subcaptionbox{Explicit Use\label{fig:tree-explicit}}{
  \begin{tikzpicture}[grow=right,sloped]
    \node[bag] {race}
    child {
      node[end, label=right:{deny}] {}
      edge from parent
      node[below] {\small\texttt{black}}
    }
    child {
      node[end, label=right:{accept}] {}
      edge from parent
      node[above] {\small\texttt{white}}
    }
    ;
  \end{tikzpicture}
}\hspace{-0.5em}%
\tikzstyle{level 1}=[level distance=1.65cm, sibling distance=1.25cm]
\tikzstyle{level 2}=[level distance=1.65cm, sibling distance=0.75cm]
\subcaptionbox{Use via proxy\label{fig:tree-redline}}{
  \begin{tikzpicture}[grow=right,sloped]
    \node[bag] {zip-code}
    child {
      node[end, label=right:{deny}] {}
      edge from parent
      node[below] {\small$b_1,b_2$}
    }
    child {
      node[end, label=right:{accept}] {}
      edge from parent
      node[above] {\small$w_1,w_2$}
    }
    ;
  \end{tikzpicture}
}\hspace{-0.5em}%
\subcaptionbox{No use\label{fig:tree-benign}}{
  \begin{tikzpicture}[grow=right,sloped]
    \node[bag] {zip-code}
    child {
      node[end, label=right:{deny}] {}
      edge from parent
      node[below] {\small$w_2,b_2$}
    }
    child {
      node[end, label=right:{accept}] {}
      edge from parent
      node[above] {\small$w_1,b_1$}
    }
    ;
  \end{tikzpicture}
}\hspace{0em}%
\tikzstyle{level 1}=[level distance=1.8cm, sibling distance=1.5cm]
\tikzstyle{level 2}=[level distance=1.6cm, sibling distance=0.75cm]
\subcaptionbox{Masked use via proxy\label{fig:tree-masked}}{
  \begin{tikzpicture}[grow=right,sloped]
    \node[bag] {zip-code}
    child {
      node[bag,align=right] {interest}
      child {
        node[end,label=right:{deny}] {}
        edge from parent
        node[below] {\small$\mathtt{high}$}
      }
      child {
        node[end,label=right:{accept}] {}
        edge from parent
        node[above] {\small$\mathtt{low}$}
      }
      edge from parent
      node[below] {\small$b_1,b_2$}
    }
    child{
      node[bag,align=right] {interest}
      child {
        node[end,label=right:{accept}] {}
        edge from parent
        node[below] {\small$\mathtt{high}$}
      }
      child {
        node[end,label=right:{deny}] {}
        edge from parent
        node[above] {\small$\mathtt{low}$}
      }
      edge from parent
      node[above] {\small$w_1,w_2$}
    };
  \end{tikzpicture}
}
 \caption{Examples of models (decision trees) used by a bank for
  accepting home loan applications.
  The bank uses race, zip-code, and customer's level of interest.
  Zip-codes $w_1$ and $w_2$ are predominantly white.
  whereas zip-codes $b_1$ and $b_2$ are predominantly black.
  Interest in loan, (\texttt{high} or \texttt{low}), is independent of
  race.}
\label{fig:proxy-usage-examples}
\end{figure*}

\begin{example}
  \label{ex:proxy1}
  (Explicit use, Fig.~\ref{fig:tree-explicit}) A bank explicitly uses
  race in order to evaluate loan eligibility.
  \cut{Figure~\ref{fig:tree-explicit} shows such an algorithm.}
\end{example}

This form of explicit use of a protected information type can be
discovered by existing black-box experimentation methods that
establish causal effects between inputs and outputs (e.g., see
\cite{datta2015opacity,sunlight,qii}).

\begin{example}
  \label{ex:proxy2}
  (Inferred use, Fig.~\ref{fig:tree-redline}) Consider a situation
  where applicants' zip-code is indicative of their race.
  The bank can thus use zip-code in place of race to evaluate loan
  eligibility as in Figure ~\ref{fig:tree-redline}.
\end{example}

This example, while very similar in effect, does not use race
directly.
Instead, it infers race via associations and then uses it.
Existing methods (see \cite{feldman15disparate,fairtest}) can detect
such associations between protected classes and outcomes in
observational data.

\begin{example}
  \label{ex:proxy3}
  (No use, Fig.~\ref{fig:tree-benign}) The bank uses some uncorrelated
  selection of zip-codes to determine eligibility.
  In Figure ~\ref{fig:tree-benign}, the zip-codes $w_1,b_1$ could
  designate suburban areas that as a category are not associated with
  race.
\end{example}

In this example, even though the bank could have inferred race from
the data available, no such inference was used in loan evaluation.
As associations are commonplace, a definition of use disallowing such
benign use of associated data would be too restrictive for practical
enforcement.

\begin{example}
  \label{ex:proxy4}
  (Masked proxy use, Fig.~\ref{fig:tree-masked}) Consider a more
  insidious version of Example~\ref{ex:proxy2}.
  To mask the association between the outcome and race, the bank
  offers loans to not just the white population, but also those with
  low expressed interest in loans, the people who would be less likely
  to accept a loan were they offered one.
  Figure~\ref{fig:tree-masked} is an example of such an algorithm.
\end{example}

While there is no association between race and outcome in both
Example~\ref{ex:proxy3} and Example~\ref{ex:proxy4}, there is a key
difference between them.
In Example~\ref{ex:proxy4}, there is an intermediate computation based
on zip-codes that is a predictor for race, and this predictor is used
to make the decision, and therefore is a case of proxy use.
In contrast, in Example~\ref{ex:proxy3}, the intermediate computation
based on zip-code is uncorrelated with race.
Distinguishing between these examples by measuring associations using
black box techniques is non-trivial.
Instead, we leverage white-box access to the code of the classifier to
identify the intermediate computation that serves as a proxy for race.
Precisely identifying the particular proxy used also aids the
normative decision of whether the proxy use is appropriate in this
setting.

\subsection{Notation and Preliminaries}\label{sec:notation}
We assume individuals are drawn from a population distribution
$\population$, in which our definitions are parametric.
Random variables $W$, $X$, $Y$, $Z$, $\ldots$ are functions over
$\population$, and the notation $W \in \cal W$ represents that the
type of random variable is $W : \cal P \to \cal W$.
An important random variable used throughout the paper is $\features$,
which represents the vector of features of an individual that is
provided to a predictive model.
A predictive model is denoted by $\pairpop{\features}{\model}$, where
$\model$ is a function that operates on $\features$.
For simplicity, we assume that $\population$ is discrete, and that
models are deterministic.
Table~\ref{tab:notation} summarizes all the notation used in this
paper, in addition to the notation for programs that is introduced
later in the paper.
Though we formalize proxies in terms of distributions and random
variables, in practice we will operate on datasets.
Datasets are samples of a population and approximate that population's
distribution.
This point is further discussed in \sref{sec:discussion-distribution}.

\setlength{\belowcaptionskip}{0pt}

\begin{table}
  \centering
  \begin{tabular}{|r | p{0.75\columnwidth} |}
    \hline
    $f$                           & A function \\
    $\pairpop{\features}{\model}$ & A model, which is a function $\model$ used for prediction, operating on random variables $\features$, in population $\population$\\
    $X$                           & A random variable\\
    $p$                           & A program \\
    $\pairpop{\features}{p}$      & A syntactic model, which is a program $p$, operating on random variables $\features$\\
    $\subst{p_1}{X}{p_2}$         & A substitution of $p_1$ in place of $X$ in $ p_2 $\\
    $\features$                   & A sequence of random variables \\
    \hline
\end{tabular}
\caption{Summary of notation used in the paper
\label{tab:notation}
}
\end{table}

\setlength{\belowcaptionskip}{-10pt}
\subsubsection{Proxies}
\label{sec:defn:proxy}

A \emph{perfect proxy} for a random variable $Z$ is a random variable $X$ that
is perfectly correlated with $Z$.  Informally, if $X$ is a proxy of $Z$, then
$X$ or $Z$ can be interchangeably used in any computation over the same
distribution. One way to state this is to require that $\Pr(X = Z) = 1$, i.e.
$X$ and $Z$ are equal on the distribution. However, we require our definition
of proxy to be invariant under renaming. For example, if $X$ is $0$ whenever
$Z$ is $1$ and vice versa, we should still identify $X$ to be a proxy for $Z$.
In order to achieve invariance under renaming, our definition only requires the
existence of mappings between $X$ and $Z$, instead of equality.

\begin{defn}[Perfect Proxy]
  \label{def:perfect-proxy}
  A random variable $X \in \cal X$ is a \emph{perfect proxy for
    $Z \in \cal Z$} if there exist functions
  $f : {\cal X} \to {\cal Z}, g : \cal{Z} \to \cal{X}$, such that
  $\Pr(Z = f(X)) = \Pr(g(Z) = X) = 1$.
\end{defn}

While this notion of a proxy is too strong in practice, it is useful
as a starting point to explain the key ideas in our definition of
proxy use.
This definition captures two key properties of proxies,
\emph{equivalence} and \emph{invariance under renaming}.

\paragraph{Equivalence} Definition~\ref{def:perfect-proxy} captures
the property that proxies admit predictors in both directions: it is
possible to construct a predictor of $X$ from $Z$, and vice versa.
This condition is required to ensure that our definition of proxy only
identifies the part of the input that corresponds to the protected
attribute and not the input attribute as a whole.
For example, if only the final digit of a zip code is a proxy for
race, the entirety of the zip code will not be identified as a proxy
even though it admits a predictor in one direction.
Only if the final digit is used, that use will be identified as proxy
use.

The equivalence criterion distinguishes benign use of
associated information from proxy use as illustrated in the next
example.
For machine learning in particular, this is an important pragmatic requirement; given enough input
features one can expect any protected class to be predictable from the
set of inputs.
In such cases, the input features taken together are a strong
associate in one direction, and prohibiting such one-sided associates from being
used would rule out most machine learnt models.

\begin{example}
  Recall that in Figure~\ref{fig:proxy-usage-examples}, zip-codes
  $w_1, w_2$ indicate white populations and $b_1,b_2$ indicate black
  populations.
  Consider Example~\ref{ex:proxy3} (No use), where zip-code is an
  influential input to the program that determines loan offers, using
  the criterion \texttt{zip-code} $\in \set{w_1, b_1}$.
  According to Definition~\ref{def:perfect-proxy}, neither race nor
  this criterion are proxies, because race does not predict zip-code
  (or specifically the value of the predicate \texttt{zip-code}
  $\in \set{w_1, b_1}$).
  However, if Definition~\ref{def:perfect-proxy} were to allow
  one-sided associations, then \texttt{zip-code} would be a proxy
  because it can predict race.
  This would have the unfortunate effect of implying that the benign
  application in Example~\ref{ex:proxy3} has proxy use of race.
\end{example}

\paragraph{Invariance under renaming} This definition of a proxy is
invariant under renaming of the values of a proxy.
Suppose that a random variable evaluates to $1$ when the protected
information type is $0$ and vice versa, then this definition still
identifies the random variable as a proxy.

\subsubsection{Influence}
\label{sec:defn:influence}

Our definition of influence aims to capture the presence of a causal
dependence between a variable and the output of a function.
Intuitively, a variable $x$ is influential on $f$ if it is possible to
change the value of $f$ by changing $x$ while keeping the other input
variables fixed.

\begin{defn}
  For a function $f(x, y)$, $x$ is influential if and only if there
  exists values $x_1$, $x_2$, $y$, such that
  $f(x_1, y) \not= f(x_2, y)$.
\end{defn}

In Figure~\ref{fig:tree-explicit}, race is an influential input of the
system, as just changing race while keeping all other inputs fixed
changes the prediction.
Influence, as defined here, is identical to the notion of
interference\cite{goguen1982security} used in the information flow
literature.

\subsection{Definition}\label{sec:definition}

We use an abstract framework of program syntax to reason about
programs without specifying a particular language to ensure that our
definition remains general.
Our definition relies on syntax to reason about decompositions of
programs into intermediate computations, which can then be identified
as instances of proxy use using the concepts described above.

\paragraph{Program decomposition}
We assume that models are represented by programs.
For a set of random variables $\features$, $\pairpop{\features}{p}$
denotes the assumption that $p$ will run on the variables in
$\features$.
Programs are given meaning by a denotation function
$\denote{\cdot}_\features$ that maps programs to functions.
If $\pairpop{\features}{p}$, then $\denote{p}$ is a function on
variables in $\features$, and $\denote{p}(\features)$ represents the
random variable of the outcome of $p$, when evaluated on the input
random variables $\features$.
Programs support substitution of free variables with other programs,
denoted by $\subst{p_1}{X}{p_2}$, such that if $p_1$ and $p_2$
programs that run on the variables $\features$ and $\features, X$,
respectively, then $\subst{p_1}{X}{p_2}$ is a program that operates on
$\features$.

A decomposition of program $p$ is a way of rewriting $p$ as two
programs $p_1$ and $p_2$ that can be combined via substitution to
yield the original program.

\begin{defn}[Decomposition]
  \label{def:decomposition}
  Given a program $p$, a decomposition $(p_1, X, p_2)$ consists of two
  programs $p_1$, $p_2$, and a fresh variable $X$, such that
  $p = \subst{p_1}{X}{p_2}$.
\end{defn}

For the purposes of our proxy use definition we view the first
component $p_1$ as the intermediate computation suspected of proxy
use, and $p_2$ as the rest of the computation that takes in $p_1$ as
an input.

\begin{defn}[Influential Decomposition]
  Given a program $p$, a decomposition $(p_1, X, p_2)$ is influential
  iff $X$ is influential in $p_2$.
\end{defn}

\paragraph{Main definition}

\begin{defn}[Proxy Use]
  \label{def:proxy-usage}
  A program $\pairpop{\features}{p}$ has proxy use of $Z$ if there
  exists an influential decomposition $(p_1, X, p_2)$ of
  $\pairpop{\features}{p}$, and $\denote{p_1}(\features)$ is a proxy
  for $Z$.
\end{defn}

\begin{example}
  In Figure~\ref{fig:tree-masked}, this definition would identify
  proxy use using the decomposition $(p_1, U, p_2)$, where $p_2$ is
  the entire tree, but with the condition
  $(a_1,a_2 \in \text{zip-code})$ replaced by the variable $U$.
  In this example, $U$ is influential in $p_2$, since changing the
  value of $U$ changes the outcome.
  Also, we assumed that the condition $(b_1,b_2 \in \text{zip-code})$
  is a perfect predictor for race, and is therefore a proxy for race.
  Therefore, according to our definition of proxy use, the model in
  \ref{fig:tree-masked} has proxy use of race.
\end{example}

\subsection{A Quantitative Relaxation}
\label{sec:defn:quant}

Definition~\ref{def:proxy-usage} is too strong in one sense and too
weak in another.
It requires that intermediate computations be perfectly correlated
with a protected class, and that there exists \emph{some} input,
however improbable, in which the result of the intermediate
computation is relevant to the model.
For practical purposes, we would like to capture imperfect proxies
that are strongly associated with an attribute, but only those whose
influence on the final model is appreciable.
To relax the requirement of perfect proxies and non-zero influence, we
quantify these two notions to provide a parameterized definition.
Recognizing that neither perfect non-discrimination nor perfect
utility are practical, the quantitative definition provides a means
for navigating non-discrimination vs.
utility tradeoffs.

\paragraph{$\epsilon$-proxies}

We wish to measure how strongly a random variable $X$ is a proxy for a
random variable $Z$.
Recall the two key requirements from the earlier definition of a
proxy: (i) the association needs to be capture equivalence and measure association in both directions, and (ii) the
association needs to be invariant under renaming of the random
variables.
The \emph{variation of information metric}
$d_{\text{var}}(X, Z) = H(X | Z) + H(Z | X)$~\cite{cover2012elements}
is one measure that satisfies these two requirements.
The first component in the metric, the conditional entropy of $X$
given $Z$, $H(X|Z)$, measures how well $X$ can be predicted from $Z$,
and $H(Z|X)$ measures how well $Z$ can be predicted from $X$, thus
satisfying the requirement for the metric measuring association in both directions.
Additionally, one can show that conditional entropies are invariant
under renaming, thus satisfying our second criteria.
To obtain a normalized measure in $[0, 1]$, we choose
$1 - \frac{d_{\text{var}}(X, Z)}{H(X, Z)}$ as our measure of
association, where the measure being $1$ implies perfect proxies, and
$0$ implies statistical independence.
Interestingly, this measure is identical to normalized mutual
information~\cite{cover2012elements}, a standard measure that has also
been used in prior work in identifying associations in outcomes of
machine learning models~\cite{fairtest}.

\begin{defn}[Proxy Association]
  Given two random variables $X$ and $Z$, the strength of a proxy is
  given by normalized mutual information,
  \begin{equation*}
    d(X, Z) \defeq 1 - \frac{H(X | Z) +
      H(Z | X)}{H(X, Z)}
  \end{equation*}
  where $X$ is defined to be an $\epsilon$-proxy for $Z$ if
  $d(X, Z) \geq \epsilon$.
\end{defn}

\paragraph{$\delta$-influential decomposition}

Recall that for a decomposition $(p_1, X, p_2)$, in the qualitative
sense, influence is interference which implies that there exists $x$,
$x_1$, $x_2$, such that
$\denote{p_2}(x, x_1) \not= \denote{p_2}(x, x_2)$. 
Here $x_1$, $x_2$ are values of $p_1$, that for a given $x$, change
the outcome of $p_2$.
However, this definition is too strong as it requires only a single
pair of values $x_1$, $x_2$ to show that the outcome can be changed by
$p_1$ alone.
To measure influence, we quantify interference by using Quantitative
Input Influence (QII), a causal measure of input influence introduced
in \cite{qii}.
In our context, for a decomposition $(p_1, X, p_2)$, the influence of
$p_1$ on $p_2$ is given by:
\begin{equation*}
  \iota(p_1, p_2) \defeq
  \mathbb{E}_{\features,\features' \samp
    \population}\prob{\denote{p_2}(\features, \denote{p_1}(\features))
    \not= \denote{p_2}(\features, \denote{p_1}(\features'))}.
\end{equation*}
Intuitively, this quantity measures the likelihood of finding randomly
chosen values of the output of $p_1$ that would change the outcome of
$p_2$.
Note that this general definition allows for probabilistic models
though in this work we only evaluate our methods on deterministic
models.

\begin{defn}[Decomposition Influence]
  Given a decomposition $(p_1, X, p_2)$, the influence of the
  decomposition is given by the QII of $X$ on $p_2$.
  A decomposition $(p_1, X, p_2)$ is defined to be
  $\delta$-influential if $\iota(p_1, p_2) > \delta$.
\end{defn}

\paragraph{$(\epsilon,\delta)$-proxy use}

Now that we have quantitative versions of the primitives used in
Definition~\ref{def:proxy-usage}, we are in a position to define
quantitative proxy use (Definition~\ref{def:quant-usage}).
The structure of this definition is the same as before, with
quantitative measures substituted in for the qualitative assertions
used in Definition~\ref{def:proxy-usage}.

\begin{defn}[$(\epsilon,\delta)$-proxy use]
\label{def:quant-usage}
A program $\pairpop{\features}{p}$ has $(\epsilon,\delta)$-proxy use
of random variable $Z$ iff there exists a $\delta$-influential
decomposition $(p_1, X, p_2)$, such that $\denote{p}(\features)$ is an
$\epsilon$-proxy for $Z$.
\end{defn}

This definition is a strict relaxation of
Definition~\ref{def:proxy-usage}, which reduces to $(1, 0)$-proxy use.

\subsection{Axiomatic Basis for Definition}
\label{sec:defn:impossibility}

We now motivate our definitional choices by reasoning about a natural
set of properties that a notion of proxy use should satisfy.
We first prove an important impossibility result that shows that no
definition of proxy use can satisfy four natural semantic properties
of proxy use.
The central reason behind the impossibility result is that under a
purely semantic notion of function composition, the causal effect of a
proxy can be made to disappear.
Therefore, we choose a syntactic notion of function composition for
the definition of proxy use presented above.
The syntactic definition of proxy use is characterized by syntactic
properties which map very closely to the semantic properties.

\begin{property}
  \label{axm:explicit-sem}
  \emph{(Explicit Use)} If $Z$ is an influential input of the model
  $\pairpopbrace{\features,Z}{\model}$, then
  $\pairpopbrace{\features,Z}{\model}$ has proxy use of $Z$.
\end{property}
This property identifies the simplest case of proxy use: if an input
to the model is influential, then the model exhibits proxy use of that
input.

\begin{property}
  \label{axm:prep-sem}
  \emph{(Preprocessing)} If a model
  $\pairpopbrace{\features,X}{\model}$ has proxy use of random
  variable $Z$, then for any function $f$ such that
  $\prob{f(\features) = X} = 1$, let
  $\model'(x) \defeq \model(x, f(x))$.
  Then, $\pairpop{\features}{\model'}$ has proxy use of $Z$.
\end{property}
This property covers the essence of proxy use where instead of being
provided a protected information type explicitly, the program uses a
strong predictor for it instead.
This property states that models that use inputs explicitly and via
proxies should not be differentiated under a reasonable theory of
proxy use.

\begin{property}
  \label{axm:dummy-sem}
  \emph{(Dummy)} Given $\pairpop{\features}{\model}$, define $\model'$
  such that for all $x, x'$, $\model'(x, x') \defeq \model(x)$, then
  $\pairpop{\features}{\model}$ has proxy use for some $Z$ iff
  $\pairpop{\{\features, X\}}{\model'}$ has proxy use of $Z$.
\end{property}
This property states that the addition of an input to a model that is
not influential, i.e., has no effect on the outcomes of the model, has
no bearing on whether a program has proxy use or not.
This property is an important sanity check that ensures that models
aren't implicated by the inclusion of inputs that they do not use.

\begin{property}
  \label{axm:indep-sem}
  \emph{(Independence)} If $\features$ is independent of $Z$ in
  $\population$, then $\pairpop{\features}{\model}$ does not have proxy
  use of $Z$.
\end{property}
Independence between the protected information type and the inputs
ensures that the model cannot infer the protected information type for
the population $\population$.
This property captures the intuition that if the model cannot infer
the protected information type then it cannot possibly use it.

While all of these properties seem intuitively desirable, it turns out
that these properties can not be achieved simultaneously.

\begin{thm}\label{thm:sem-impossibility}\label{THM:SEM-IMPOSSIBILITY}
  No definition of proxy use can satisfy Properties
  \ref{axm:explicit-sem}-\ref{axm:indep-sem} simultaneously.
\end{thm}

See our companion paper~\cite[Appendix A]{useprivacy} for a proof of
the impossibility result and a discussion.
The key intuition behind the result is that Property
\ref{axm:prep-sem} requires proxy use to be preserved when an input is
replaced with a function that predicts that input via composition.
However, with a purely semantic notion of function composition, after
replacement, the proxy may get canceled out.
To overcome this impossibility result, we choose a more syntactic
notion of function composition, which is tied to how the function is
represented as a program, and looks for evidence of proxy use within
the representation.

We now proceed to the axiomatic justification of our definition of
proxy use.
As in our attempt to formalize a semantic definition, we base our
definition on a set of natural properties given below.
These are syntactic versions of their semantic counterparts defined
earlier.

\begin{property}
  \label{axm:syn-explicit}
  \emph{(Syntactic Explicit Use)} If $X$ is a proxy of $Z$, and $X$ is
  an influential input of $\pairpopbrace{\features, X}{p}$, then
  $\pairpopbrace{\features, X}{p}$ has proxy use.
\end{property}

\begin{property}
  \label{axm:syn-preprocessing}
  \emph{(Syntactic Preprocessing)} If
  $\pairpopbrace{\features, X}{p_1}$ has proxy use of $Z$, then for
  any $p_2$ such that $\prob{\denote{p_2}(\features) = X} = 1$,
  $\pairpop{\features}{\subst{p_2}{X}{p_1}}$ has proxy use of $Z$.
\end{property}

\begin{property}
  \label{axm:syn-dummy}
  \emph{(Syntactic Dummy)} Given a program $\pairpop{\features}{p}$,
  $\pairpop{\features}{p}$ has proxy use for some $Z$ iff
  $\pairpopbrace{\features, X}{p}$ has proxy use of $Z$.
\end{property}

\begin{property}
  \label{axm:syn-independence}
  \emph{(Syntactic Independence)} If $\features$ is independent of
  $Z$, then $\pairpop{\features}{p}$ does not have proxy use of $Z$.
\end{property}

Properties \ref{axm:syn-explicit} and \ref{axm:syn-preprocessing}
together characterize a complete inductive definition, where the
induction is over the structure of the program.
Suppose we can decompose programs $p$ into $(p_1, X, p_2)$ such that
$p = \subst{p_1}{X}{p_2}$.
Now if $X$, which is the output of $p_1$, is a proxy for $Z$ and is
influential in $p_2$, then by Property~\ref{axm:syn-explicit}, $p_2$
has proxy use.
Further, since $p = \subst{p_1}{X}{p_2}$, by
Property~\ref{axm:syn-preprocessing}, $p$ has proxy use.
This inductive definition where we use Property~\ref{axm:syn-explicit}
as the base case and Property~\ref{axm:syn-preprocessing} for the
induction step, precisely characterizes
Definition~\ref{def:proxy-usage}.
Additionally, it can be shown that Definition~\ref{def:proxy-usage}
also satisfies Properties~\ref{axm:syn-dummy}
and~\ref{axm:syn-independence}.
Essentially, by relaxing our notion of function composition to a
syntactic one, we obtain a practical definition of proxy use
characterized by the natural axioms above.

\section{Detection and Repair of \ProxyUse}
\label{sect:detection-and-repair}

In this section, we summarize algorithms for 1) identifying \proxyuse
of specified variables in a given machine-learning model and 2)
repairing those models so that the \proxyuse is removed. 
Details of these algorithms are presented in the companion
paper~\cite{useprivacy}. 
There the reader can also find proofs of the theorems noted in this
section as well as various optimizations that are part of our
implementations.

\subsection{Environment Model}
\label{sec:detecting:model}

The environment in which our detection algorithm operates is comprised
of a data processor, a dataset that has been partitioned into analysis
and validation subsets, and a machine learning model trained over the
analysis subset.
The data processor is responsible for determining whether the model
contains any instances of proxy use, and works cooperatively with the
algorithm towards this goal.
In other words, we assume that the data processor does not act to
evade the detection algorithm, and provides accurate information.
As such, our algorithm assumes access both to the text of the program
that computes the model, as well as the analysis and validation data
used to build it.
Additionally, we assume that attributes indicating protected classes
we wish to detect proxies of are also part of the validation data.
We discuss this point further in Section~\ref{sec:discussion}.

\subsection{Models as expression programs}
Our techniques are not tied to any particular language, and the key
ideas behind them apply generally. 
For our implementation work we focused on a simple \emph{expression}
(functional) language that is rich enough to support commonly-used
models such as decision trees, linear and logistic regression, Naive
Bayes, and Bayesian rule lists.
Programs denote functions that evaluate arithmetic expressions, which
are constructed from real numbers, variables, common arithmetic
operations, and if-then-else constructs.

Boolean expressions, which are used as conditions in if-then-else
expressions, are constructed from the usual connectives and relational
operations.
Finally, we use $\lambda$-notation for functions, i.e., $\lambda x .
e$ denotes a function over $x$ which evaluates $e$ after replacing all
instances of $x$ with its argument.
Details of this language and how machine learning models such as
linear models, decision trees, and random forests are translated to
this expression language are discussed in the companion
paper~\cite[B.2]{useprivacy}. 
The consequences of the choice of language and decomposition in that
language are further discussed in Section~\ref{sec:discussion}.

\paragraph{Distributed proxies}
Our use of program decomposition provides for partial handling of
\emph{distributed representations}, the idea that concepts can be
distributed among multiple entities.
In our case, influence and association of a protected class can be
distributed among multiple program points.
First, substitution (denoted by $[p_1/X]p_2$) can be defined to
replace \emph{all} instances of variable $X$ in $p_2$ with the program
$p_1$.
If there are multiple instances of $ X $ in $ p_2 $, they are still
describing a single decomposition and thus the multiple instances of
$ p_2 $ in $p_1$ are viewed as a single proxy.
Further, implementations of substitution can be (and is in our
implementation) associativity-aware: programs like $x_1 + x_2 + x_3$
can be equivalent regardless of the order of the expressions in that
they can be decomposed in exactly the same set of ways.
If a proxy is distributed among $x_1$ and $x_3$, it will still be
considered by our methods because $ x_1 + (x_2 + x_3) $ is equivalent
to $ (x_1 + x_3) + x_2 $, and the sub-expression $ x_1 + x_3 $ is part
of a valid decomposition.

\subsection{Analyzing Proxy Use} \label{sec:detecting-analyzing}

\begin{algorithm}[t]
  \caption{Detection for expression programs.
    \label{alg:detect-generic-informal}}
  \begin{algorithmic}
    \Require association (\assocmeasure{}), influence(\infmeasure{}) measures
    \Procedure{ProxyDetect}{$p, \features, Z, \epsilon, \delta$}
    \State $P \gets \varnothing$
    \For{each subprogram  $p_1$ appearing in $p$}
      \For{each program $p_2$ such that $[p_2/u]p_1 = p$}
        \If{$\iota(p_1, p_2) \geq \delta \land d(\llbracket p_1 \rrbracket(\features), Z) \geq \epsilon$}
          \State $P \gets P \cup \{(p_1, p_2)\}$
        \EndIf
      \EndFor
    \EndFor
    \State \Return $P$
    \EndProcedure
\end{algorithmic}
\end{algorithm}

Algorithm~\ref{alg:detect-generic-informal} describes a general
technique for detecting \edproxyuse in expression programs.
In addition to the parameters and model expression, it operations on a
description of the distribution governing the feature variables $X$
and $Z$.
In practice this will nearly always consist of an empirical sample,
but for the sake of presentation we simplify here by assuming the
distribution is explicitly given.
In Section~\ref{sec:estimates}, we describe how the algorithm can
produce estimates from empirical samples.

The algorithm proceeds by enumerating sub-expressions of the given
program.
For each sub-expression $e$ appearing in $p$, $\Call{ProxyDetect}{}$
computes the set of positions at which $e$ appears. 
If $e$ occurs multiple times, we consider all possible subsets of
occurrences as potential decompositions\footnote{This occurs often in
  decision forests.}.
It then iterates over all combinations of these positions, and creates
a decomposition for each one to test for \edproxyuse.
Whenever the provided thresholds are exceeded, the decomposition is
added to the return set.
This proceeds until there are no more subterms to consider.
While not efficient in the worst-case, this approach is both sound and
complete with respect to Definition~\ref{def:quant-usage}. 
It is important to mention, however, that our definitions are not
meant to capture all types of indirect discrimination; completeness
and soundness here is therefore only in relation to an incomplete
definition.

\begin{thm}[Detection soundness]
  \label{thm:detect-sound} Any decomposition $(p_1, p_2) $ returned by
  $ \Call{ProxyDetect}{p, \features, \epsilon, \delta}$ is a
  decomposition of the input program $ p $ and had to pass the
  $ \epsilon,\delta $ thresholds, hence is a \edproxyuse.
\end{thm}

\begin{thm}[Detection completeness]\label{thm:detect-complete}
  Every decomposition which could be a \edproxyuse is enumerated by
  the algorithm.
  Thus, if $(p_1, p_2)$ is a decomposition of $p$ with
  $\iota(p_1, p_2) \geq d$ and
  $d(\llbracket p_1 \rrbracket(\features), Z) \geq \epsilon$, it will
  be returned by $\Call{ProxyDetect}{p, \features, \epsilon, \delta}$.
\end{thm}

Our detection algorithm considers single expressions in its
decomposition. 
Sometimes a large number of syntactically different proxies with weak
influence might collectively have high influence. 
A stronger notion of program decomposition that allows a collection of
multiple different expressions to be considered a proxy would identify
such a case of proxy use but will have to search over a larger space
of expressions. 
Exploring this tradeoff between scalability and richer proxies is an
important topic for future work.

\subsubsection{Estimating influence and association}
\label{sec:estimates}

It is rarely the case that one has access to the precise distribution
from which data is drawn. 
Instead, a finite sample must be used as a surrogate when reasoning
about random variables. 
We describe how the two primary quantities used in
$\Call{ProxyDetect}{}$, influence and association, are estimated from
such a sample. 
The use of a sample in place of a distribution is discussed in
\sref{sec:discussion-distribution}.

\paragraph{Quantitative decomposition influence}

Given a decomposition $(p_1, u, p_2)$ of $p$, Algorithm
$\Call{ProxyDetect}{}$ first calculates the influence of $p_1$ on
$p_2$'s output to ensure that the potential proxy quantity is relevant
to the model's output. 
Recall that this influence is defined as:
\begin{equation*}
  \iota(p_1, p_2) \defeq \mathbb{E}_{X, X' \samp \population}
  \prob{
    \denote{p_2}\paren{\features, \denote{p_1}{\features}} \ne
    \denote{p_2}\paren{\features, \denote{p_1}{\features'}}
  }
\end{equation*}
Assuming deterministic models and given a dataset $\mathcal{D}$ drawn
from $\population$, we estimate this expectation by aggregating over
the rows:
\begin{equation*}
  \hat{\iota}(p_1, p_2)
  \defeq
  \frac{1}{|\mathcal{D}|^2}
  \sum_{\mathbf{x} \in \mathcal{D}}\sum_{\mathbf{x}' \in \mathcal{D}}
  \mathbbm{1}\paren{
    \denote{p_2}\paren{\mathbf{x}, \denote{p_1}{\mathbf{x}}} \ne
    \denote{p_2}\paren{\mathbf{x}, \denote{p_1}{\mathbf{x}'}}
  }
\end{equation*}

The quadratic cost of this computation makes it infeasible when
$\mathcal{D}$ is large, so in practice we take a \emph{sample} from
$\mathcal{D} \times \mathcal{D}$.
By Hoeffding's inequality~\cite{Hoeffding:1963}, we select the
subsample size $n$ to be at least $\log(2/\beta)/2\alpha^2$ to ensure
that the probability of the error
$\hat{\iota}(p_1, p_2) - \iota(p_1, p_2)$ being greater than $\beta$
is bounded by $\alpha$.
An additional optimization follows if we introduce a notion of
\emph{reachability} for subexpressions.
An input $ X $ reaches a sub-expression $p_1$ inside $ p $ if the
evaluation of $ p $ on $ X $ requires evaluating $ p_1 $.
Using reachability, we improve the computation of influence by
realizing that if an input $ X $ does not reach $ p_1 $, there is no
value we can replace $ p_1 $ with that will change the outcome of
$ p $ evaluated on $ X $.
\cut{
  \begin{align*}
    \iota(p_1, p_2)
    &\defeq \expectsub{\features, \features' \samp \population}{\indicator{\denote{p}{\features} \neq \denote{p_2}\paren{\features, \denote{p_1}{\features'}}}} \\
    &= \prob{p_1 \text{ reached}}\cdot \expectsub{\features \samp \population|p_1 \text{ reached}}{\cdots} \\
    & \;\;\;\; + \prob{p_1 \text{ not reached}} \cdot \expectsub{\features \samp \population|p_1 \text{ not reached}}{\cdots} \\
    & = \prob{p_1 \text{ reached}}\cdot \\
    & \;\;\;\; \expectsub{\features \samp \population|p_1 \text{ reached}}{\expectsub{\features' \samp \population}{\indicator{p(\features) \neq \denote{p_2}\paren{\features, \denote{p_1}{\features'}}}}}
  \end{align*}
}
When estimating influence, we take advantage of the optimization by
conditioning our sampling on the reachability of the decomposed
subexpression $p_1$.

\paragraph{Association}

As discussed in Section~\ref{sec:defn}, we use mutual information to
measure the association between the output of a subprogram and $Z$. 
This quantity can be estimated from a sample in time
$O(|\mathcal{D}| + k|\mathcal{Z}|)$, where $k$ is the number of
elements in the range of $p_1$~\cite{Meila2003}. 
For each $(\mathbf{x}, z)$ in the dataset, the procedure computes
$p_1$ on $\mathbf{x}$ and builds a \emph{contingency table} indexed by
$(\llbracket p_1\rrbracket(\mathbf{x}), z)$. 
The contingency table is used to compute the required conditional
entropies. 
A particular concern while estimating associations is associations
appearing by random chance on a particular sample. 
In the companion paper~\cite[Appendix B.3]{useprivacy} we discuss how
to mitigate the reporting of such spurious associations.

\subsection{Removing Proxy Use Violations} \label{sect:repair}

\begin{algorithm}[t]
\caption{Witness-driven repair.}
\begin{algorithmic}
\Require association (\assocmeasure{}), influence (\infmeasure{}), utility (\utilmeasure{}) measures, oracle ($\oracle{}$)
\Procedure{Repair}{$p, \features, Z, \epsilon, \delta$}
\State $ P  \gets \set{d \in \Call{ProxyDetect}{p,\features,Z,\epsilon, \delta} : \text{ not } \oracle(d)} $
\If{ $ P \neq \emptyset $ }
\State $(p_1,p_2) \gets \text{element of } P $
\State $p' \gets \Call{ProxyRepair}{p, (p_1,p_2), \features,Z,\epsilon, \delta} $
\State \Return $\Call{Repair}{p',\features, Z, \epsilon, \delta} $
\Else
\State \Return $p$
\EndIf
\EndProcedure
\end{algorithmic}
\label{alg:repair-informal}
\end{algorithm}

\begin{algorithm}[t]
\caption{Local Repair.}
\begin{algorithmic}[1]
\Require association (\assocmeasure{}), influence (\infmeasure{}), utility (\utilmeasure{}) measures
\Procedure{ProxyRepair}{$p, (p_1,p_2), \features, Z, \epsilon, \delta$}
\State $R \gets \set{}$
\For{each subprogram $p_1'$ of $p_1$}\label{line:repair-local}
\State $r^* \gets$~Optimal constant for replacing $p_1'$ \label{line:repair-optimal}
\State $(p_1'',p_2'') \gets (p_1,p_2) $ with $ r^* $ subst. for $ p_1' $ \label{line:repair-subst}
\If{$\iota(p_1'', p_2'') \leq \delta \lor d(\llbracket p_1'' \rrbracket(\features), Z) \leq \epsilon$} \label{line:repair-violation}
\State $ R \gets R \cup \subst{u}{r^*}{p_2'}$
\EndIf
\EndFor
\State \Return $\argmax_{p^*\in R} \utilmeasure\paren{p^*}$ \label{line:repair-optimal2}
\EndProcedure
\end{algorithmic}
\label{alg:repair-local-informal}
\end{algorithm}

\newcommand{\lref}[1]{Line~\ref{#1}}

Our approach for removing proxy use violations has two parts: first
($\Call{Repair}{}$, Algorithm~\ref{alg:repair-informal}) is the
iterative discovery of proxy uses via the $\Call{ProxyDetect}{}$
procedure described in the previous section and second
($\Call{ProxyRepair}{}$, Algorithm~\ref{alg:repair-local-informal}) is
the repair of the ones found by the oracle to be violations. 
Our repair procedures operate on the expression language, so they can
be applied to any model that can be written in the language. 
Further, our violation repair algorithm does not require knowledge of
the training algorithm that produced the model. 
The witnesses of proxy use localize where in the program violations
occur. 
To repair a violation we search through expressions \emph{local} to
the violation, replacing the one which has the least impact on the
accuracy of the model that at the same time reduces the association or
influence of the violation to below the $ (\epsilon,\delta) $
threshold.

At the core of our violation repair algorithm is the simplification of
sub-expressions in a model that are found to be violations.
Simplification here means the replacement of an expression that is not
a constant with one that is.
Simplification has an impact on the model's performance hence we take
into account the goal of preserving utility of the machine learning
program we repair.
We parameterize the procedure with a measure of utility $v$ that
informs the selection of expressions and constants for simplification.
We briefly discuss options and implementations for this parameter
later in this section.

The repair procedure ($\Call{ProxyRepair}{}$) works as follows.
Given a program $ p $ and a decomposition $(p_1,p_2) $, it first finds
the best simplification to apply to $ p $ that would make
$ (p_1,p_2) $ no longer a violation.
This is done by enumerating expressions that are \emph{local} to
$ p_1 $ in $ p_2 $.
Local expressions are sub-expressions of $ p_1 $ as well as $ p_1 $
itself and if $ p_1 $ is a guard in an if-then-else expression, then
local expressions of $ p_1 $ also include that if-then-else's true and
false branches and their sub-expressions.
Each of the local expressions corresponds to a decomposition of $ p $
into the local expression $ p_1' $ and the context around it $ p_2' $.
For each of these local decompositions we discover the best constant,
in terms of utility, to replace $ p_1' $ with .
We then make the same simplification to the original decomposition
$(p_1, p_2)$, resulting in $ (p_1'',p_2'') $. 
Using this third decomposition we check whether making the
simplification would repair the original violation, collecting those
simplified programs that do.
Finally, we take the best simplification of those found to remove the
violation (\lref{line:repair-optimal2}). 
Details on how the optimal constant is selected is described in the
companion paper \cite[Appendix C.1]{useprivacy}.

Two important things to note about the repair procedure.
First, there is always at least one subprogram that will fix the
violation, namely the decomposition $(p_1,p_2)$ itself.
Replacing $ p_1 $ with a constant in this case would disassociate it
from the protected class.
Secondly, the procedure produces a model that is smaller than the one
given to it as it replaces a non-constant expression with a constant.
These two let us state the following:

\begin{thm}\label{thm:repair}
  Algorithm $\Call{LocalRepair}{}$ terminates and returns a program
  that does not have any \edproxyusage violations (instances of
  \edproxyusage for which oracle returns false).
\end{thm}

\section{Evaluation}
\label{sect:eval}
In this section we empirically evaluate our definition and algorithms
on use-cases based real datasets. 
We demonstrate a detection and repair scenario in
\sref{sec:eval-scenario} and present two additional use-cases of our
theory and algorithms in \sref{sec:eval-cases}.
We describe our findings of interesting proxy uses and demonstrate how
the outputs of our detection tool would allow a normative judgment
oracle to determine the appropriateness of proxy uses.
We begin by noting some details regarding our implementation and the
datasets.

\paragraph{Models and Implementation}
Our implementation currently supports linear models, decision trees,
random forests, and rule lists. 
Note that these model types correspond to a range of commonly-used
learning algorithms such as logistic regression, support vector
machines~\cite{cortes95svm}, CART~\cite{Breiman01}, and Bayesian rule
lists~\cite{letham2015}. 
Also, these models represent a significant fraction of models used in
practice in predictive systems that operate on personal information,
ranging from advertising~\cite{chickering00advertising},
psychopathy~\cite{hare03pclr}, criminal
justice~\cite{berk14forecasts,berk16forecasting}, and actuarial
sciences~\cite{frees2014asbook,gepp16insurance}. 
In this paper we evaluate our methods on decision trees while
discussion of results on other modes can be found in the companion
paper~\cite{useprivacy}.
Our prototype implementation was written in Python, and we use
scikit-learn package to train the models used in the evaluation.
Our implementation is available at
\url{https://sites.google.com/site/proxynondiscrimination}.

\subsection{Datasets}

\paragraph{Adult} The UCI Adult dataset is widely used in the privacy
and fairness literature as a benchmark for evaluating new techniques.
It contains roughly 48,000 instances consisting of demographic
information and a classification of the individual as making more or
less than \$50,000 per year, which we can interpret as a loan decision
(predicting income of greater than \$50,000 corresponds to an accepted
loan, while less is a rejection).
To maintain consistency with prior work using this
benchmark~\cite{feldman15disparate,kamishima11fairness}, we treat
gender as the protected attribute in our scenario.

\begin{figure}[t]

\begin{lstlisting}[style=customlisp]
;; This is a quite discriminatory rule
;; against feminine, but the reality
;; of Japan presently seems so.
(def-rule jobless_unmarried_fem_reject
   (((Jobless_unmarried_fem_reject ?s)
     (jobless ?s)
     (female ?s)
     (unmarried ?s))))
\end{lstlisting}

\caption{Example domain theory rule from the Japanese Credit dataset.
  This rule encodes that unemployed, unmarried females should be
  denied credit.}
\label{fig:jc-domrule}
\end{figure}

\paragraph{Japanese Credit} The Japanese Credit dataset was collected
in 1992 from examples of 125 individuals who placed consumer credit
applications.
It contains a number of demographic and financial attributes for each
individual, and is available in the UCI
repository~\cite{lichman2013uci} in two formats.
The first is contains fifteen features whose names and values have
been randomly chosen to protect privacy, as well as a binary
classification variable corresponding to an accepted or rejected
credit application.
The second format is a set of fifteen Lisp predicates over row indices
with descriptive names (purchase-item, jobless, male, female,
unmarried, problematic-region, age, deposit, monthly-payment,
num-months, num-years-in-company), and an accompanying domain theory
provided by an expert.
An example rule from the domain theory is shown in
Figure~\ref{fig:jc-domrule}, which reflects a policy of denying credit
to jobless, unmarried females; the remark on the discriminatory nature
of this rule is taken verbatim from the original file.
The row predicates describe attribute values for each individual,
whereas the domain theory is a set of rules written in Lisp that
operate over the row predicates to determine a credit decision.
We extracted a row-structured dataset from the second format, removed
redundant attributes, and treated gender as the protected attribute in
our experiments.

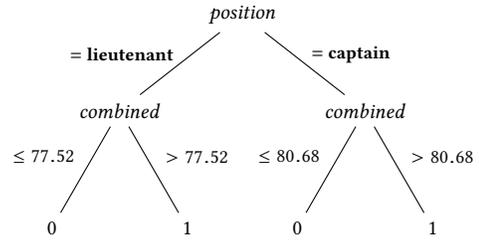
\begin{figure}[t]

\newdimen\nodeDist
\nodeDist=18mm
\centering

\begin{tikzpicture}[
    node/.style={%
      font=\small
    },
  ]

    \node [node,draw=none] (A) {$\mathit{position}$};
    \path (A) ++(-142:1.15\nodeDist) node [node,draw=none] (B) {$\mathit{combined}$};
    \path (A) ++(-38:1.15\nodeDist) node [node,draw=none] (C) {$\mathit{combined}$};
    \path (B) ++(-120:\nodeDist) node [node,draw=none] (D) {0};
    \path (B) ++(-60:\nodeDist) node [node,draw=none] (E) {1};
    \path (C) ++(-120:\nodeDist) node [node,draw=none] (F) {0};
    \path (C) ++(-60:\nodeDist) node [node,draw=none] (G) {1};

    \draw (A) -- (B) node [left,pos=0.35,xshift=-1ex] {\footnotesize$=$ \textbf{lieutenant}}(A);
    \draw (A) -- (C) node [right,pos=0.35,xshift=1ex] {\footnotesize$=$ \textbf{captain}}(A);
    \draw (B) -- (D) node [left,pos=0.35,xshift=-1ex] {\footnotesize $\leq 77.52$}(A);
    \draw (B) -- (E) node [right,pos=0.35,xshift=1ex] {\footnotesize $> 77.52$}(A);
    \draw (C) -- (F) node [left,pos=0.35,xshift=-1ex] {\footnotesize $\leq 80.68$}(A);
    \draw (C) -- (G) node [right,pos=0.35,xshift=1ex] {\footnotesize $> 80.68$}(A);
\end{tikzpicture}

\caption{Decision tree used to determine promotion eligibility in the
  \emph{Ricci v. 
    DeStefano} case.}
\label{fig:ricci-tree}
\end{figure}

\paragraph{Ricci v. 
  DeStefano} This dataset comes from the U.S. 
District Court of Connecticut's decision on the Ricci v. 
DeStefano case~\cite{ricci}. 
It contains the oral, written, and combined promotion exam scores, as
well as the race (Black, White, or Hispanic) and position (Captain or
Lieutenant), of 118 New Haven firefighters. 
The fire department's policy at the time stipulated that any applicant
with a combined score of 70\% or above is eligible for promotion, and
that whenever $n$ promotions are available, applicants must be
selected from among the top $n+2$ scorers. 
However, the department decided not to promote anyone in this case
because too few minorities matched this criteria. 
A subset of the test-takers filed a reverse-discrimination lawsuit,
and the Supreme Court eventually ruled in their favor. 
We examined the latter decision rule for proxy usage, using race as
the protected feature. 
As the case questioned the treatment of all minorities alongside that
of non-minorities, we collapsed the Black and Hispanic labels into a
single minority label. 
The decision tree corresponding to the latter policy is shown in
Figure~\ref{fig:ricci-tree}.

\subsection{Detection and Repair Scenario} \label{sec:eval-scenario}

\begin{figure*}[t]
\centering\includegraphics[width=\linewidth]{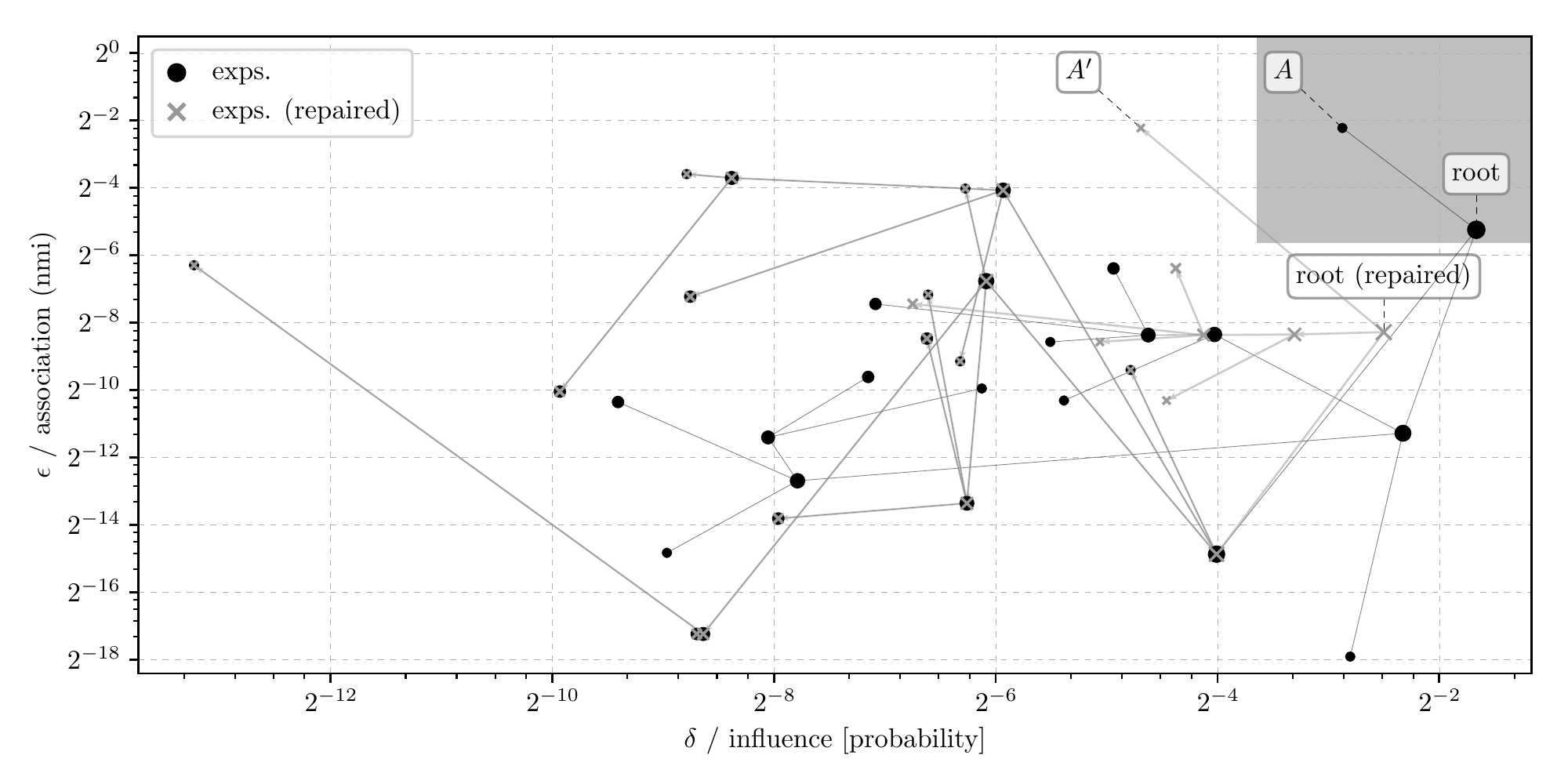}
\caption{\label{fig:plot_adult_repair} The association and influence
  of the expressions composing a decision tree trained on the UCI
  Adult dataset.
  Original tree expressions are denoted by $\bullet$ while repaired
  tree expressions are designated by $\times$.
  Dark area in the upper-left designates the thresholds used in
  repair.
  Narrow lines designate the sub-expression relationship.
  Marker size denotes the relative size of the sub-expressions
  pictured.
}
\end{figure*}

In this scenario, a bank uses income prediction data such as the UCI
Adult dataset to train a model for determining whether to accept
loans.
The bank uses the predicted income from this model to produce its
decision on a loan application (high income results in an accepted
loan, while low income results in a rejected loan).

Loan decisions are covered under the protections of the United States
Fair Lending Regulations\cite{fed06book} and gender is one of the
protected classes.
Thus the bank would like to make sure that a loan applicant's gender
is not used in its decision procedure.
They begin by first removing gender from the dataset before training
their model.
Then, using our detection procedure, they can check that in the model
they train, there are no proxy uses of gender that have high impact on
the decision.

We construct a decision tree model from this dataset as a
representative of the kind of model the bank could use, and analyze
the potential proxy uses of gender that could be present.
Figure~\ref{fig:plot_adult_repair} visualizes all of the expressions
making up the model (marked as $\bullet$), along with their
association and influence measures.
In decision trees, sub-expressions like these coincide with
decompositions in our proxy use definition; each sub-expression can be
associated with a decomposition that cuts out that sub-expression from
the tree, and leaves a variable in its place.

The point labeled $A$ in Figure~\ref{fig:plot_adult_repair} is the
predicate \texttt{relationship $\leq$ 0.5} and has significant
correlation with gender and influence (it is the predicate of the root
note of the tree).
On further examination, the relationship status in this dataset
encodes gender in most cases as \texttt{husband} and \texttt{wife} are
two of its possible values\footnote{Note that we pre-processed nominal
  features into numeric ones for our experiments.}.
This use would most likely be deemed inappropriate as modeled by the
normative oracle and thus we will remove it.
The ideal solution to this problem is to remove gender indicators from
from the \texttt{relationship} attribute but for the sake of this
demonstration, we instead use our repair algorithm.

We determine that any proxies with association and influence exceeding
the thresholds indicated by the shaded area in
Figure~\ref{fig:plot_adult_repair} are too strong; any decomposition
or sub-expression in that region is unacceptable.
This area includes the problematic predicate as well as the root of
the model, indicating that the decision procedure's outcome is itself
associated with gender to a significant enough degree.

Applying our repair procedure to this model, with the association and
influence thresholds as indicated in the figure, we produce another
tree.
This model is designated with $\times$ in
Figure~\ref{fig:plot_adult_repair}.
Note that this repaired version has no sub-expressions in the
prohibited range while a lot of the tree remains unchanged (the
$\bullet$ and $\times$ markers largely coincide).
Interestingly, the problematic predicate is still in the model, but
now at point $A'$, which has much lower influence than it had in the
un-repaired tree.
This occurred because the repair procedure did not replace the
predicate itself but instead it replaced one of the deeper predicates
(\texttt{education\_num $\leq$ 11.5}) in one of the branches of the
root node.
This replacement also reduced the association of the whole tree with
gender.

In general\footnote{Repair can improve test accuracy as it can serve
  as a regularizer.
  For some training algorithms, repair can occasionally have no train
  accuracy impact.}
repair comes with a cost of utility, or the accuracy of the repaired
model as related to the original.
The techniques we presented here and specifically the parameterized
$(\epsilon,\delta)$-proxy use definition must therefore navigate the
trade-off between fairness and utility.

\subsection{Other Use Cases}\label{sec:eval-cases}

\subsubsection{Exam score proxy in Ricci v.
  DeStefeno} We analyzed the models used in the Ricci v.
DeStefeno case to understand whether the phenomenon documented in the
case is reflected as proxy usage.
The association between race and outcomes in this data is apparently
paradoxical: an analysis of the test scores reveals a substantial
difference between minority and non-minority applicants, but an
analysis of the passing and promotion eligibility rates (as defined by
the model) shows no appreciable difference.
Recalling Figure~\ref{fig:ricci-tree}, we found that the entire tree
showed low association scores ($\epsilon=0.044$).
However, both subtrees showed a larger association ($ \epsilon=0.051 $
and $ \epsilon=0.060 $).
That is, both the lieutenant and captain decisions were more
correlated with race than the procedure as a whole.

The reason for this is simple.
For the lieutenant test, there were zero non-minorities who met the
criteria, and only three for the captain test.
This means that both subtrees signaled non-minority status when it
output a positive classification, using race as a proxy for that
particular outcome.
Note, however, that the association strength implies that the outcome
of this subtree is not a perfect proxy; while a passing
lieutenant grade perfectly predicts non-minority status, a failing one
does not.
Although the absolute association numbers may appear small, the
difference in association strength between the final output and the
intermediate computations reveals the contentious issue in the case,
and the seemingly paradoxical associations.
Without looking into the model, we would have no way of identifying
this underlying phenomenon.

\subsubsection{Purchase proxy in Japanese Credit}
As discussed previously, the domain theory for the Japanese Credit
dataset refers to gender as a critical factor in some decisions.
We trained a decision tree model on the Japanese Credit dataset after
removing gender to understand whether the learning algorithm
introduces proxy usage of gender to compensate for this missing data.
In this case, the entire model had low association with gender,
showing $(\epsilon=0.005,\delta=0.25)$-proxy usage.
The predicate used in on the root of this tree, \texttt{jobless $\leq$
  0.5}, has a much stronger association ($\epsilon=0.020$).
Joblessness could, however, be excused as a business necessity in
evaluating credit worthiness.

\section{Related Work}\label{sec:related}

\subsection{Discrimination Definitions}
The literature on use restrictions has typically focused on explicit
use of protected information types, not on proxy use (see Tschantz et
al.~\cite{tschantz12sp} for a survey and Lipton and
Regan~\cite{lipton16privacy}). 
Recent work on discovering personal data use by black-box web services
focuses mostly on explicit use of protected information types by
examining causal effects \cite{datta15pets,sunlight}; some of this
work also examines associational effects~\cite{xray,sunlight}. 
Associational effects capture some forms of proxy use but not others
as we argued in Section~\ref{sec:defn}.

A number of definitions of information use have been proposed in prior
work.
We categorize these definitions into two types: (i) associative
notions, which measure the association between inputs or outputs of
the system, and the attribute under consideration, or (ii) explicit
use notions, which identify the causal effect of the attribute under
consideration on the outcomes of a system.
Our formalization of proxy usage can be viewed as a synthesis of these
two notions where a proxy (measured by associations) has a causal
influence on the outcome (measured by causal influence).

\paragraph*{Associative Notions}

Disparate impact, a legal term introduced in Griggs v.
Duke Power Co., 1971~\cite{disparate-impact}, measures the difference
in the statistical outcomes of different groups under the protected
attribute.
This formulation of discrimination has been adopted by a number of
technical approaches to preventing
discrimination~\cite{kamishima11fairness, calders10naive,
  zemel13fair}.
Disparate impact is a special case of an association measure for
outcomes that is generalized by Tram{\`{e}}r et al.
\cite{fairtest} who provide a framework for finding conditional
associations in sub-populations.
In a different approach, Feldman et al.~\cite{feldman15disparate}
restrict the association between the inputs that are provided to the
model with the protected attribute, as such a restriction is
guaranteed to restrict an association with the outcome for any model.
We argue in \S\ref{sect:intro} that the presence of association with
outcomes is not necessary for the presence of proxy usage, while the
presence of association with inputs is not sufficient for the presence
proxy usage.
In this sense, our definition of proxy usage is not subsumed by any of
these prior works.

\paragraph*{Explicit Use Notions}

An alternate approach to defining information use is by identifying an
explicit causal influence of a protected attribute on the
outcome~\cite{sunlight,datta2015opacity}.
Another explicit use notion is Quantitative Input Influence -- a
family of causal influence measures~\cite{qii} that we use in this
paper to quantify the influence of a sub-computation.
The explicit use approach requires the protected attribute to be an
actual input to the model under scrutiny.
Thus, it does not account for proxy usage in machine learning
applications.
Dwork et al.~\cite{dwork11fairness} define fairness in terms of
Lipschitz continuity.
Their definition states that similar people are treated similarly,
which ensures that irrelevant inputs have no explicit causal influence
on the outcomes.
Irrelevant inputs are encoded in the choice of a domain specific
distance metric.
In principle, appropriate distance metrics could rule out proxy usage.
However, they do not provide a method for constructing such distance
metrics.
They leave it to future work.
The follow-up work of Zemel et al.~\cite{zemel13fair} mentioned above
ensures that there is no disparate impact but does not eliminate proxy
use.

\paragraph*{Causal Indirect Use}

Adler et al.~\cite{adler2016auditing} describe a method for estimating
the indirect influence of a protected class on a model's outcome by
computing that model's accuracy on a dataset in which proxies of the
protected class have been obscured. 
They argue that the difference between this accuracy and accuracy on
the un-obscured data is a measure of the protected class's influence
and can thus determine whether it is a cause of disparate outcomes.
Their technique does not rely on white-box access to the models but
assumes that proxies-class relationship can be learned by a given set
of algorithms. 
Our setting and assumptions differ in that we make no assumptions
about the proxy-class relationship though we require white-box access.
We also provide repair algorithms that can strip proxy use from
previously learnt models.

Kilbertus et al.~\cite{kilbertus2017avoiding} follow Pearl's work on
the discrimination~\cite{CausalityBook} by describing indirect/proxy
discrimination in terms of causal graphs. 
They also discuss algorithms for avoiding such discrimination in some
circumstances. 
Our work does not rely on presence of a causal graph to specify the
causal relationships between features.

\subsection{Detection and Repair Methods}

Our detection algorithm operates with white-box access to the
prediction model which is a stronger access assumption.

\paragraph{Access to observational data} Detection techniques working
under an associative use definition~\cite{fairtest,
  feldman15disparate} usually only require access to observational
data about the behavior of the system.

\paragraph{Access to black-box experimental data} Detection techniques
working under an explicit use definition of information
use~\cite{sunlight,datta15pets} typically require experimental access
to the system. 
This access allows the analyst to control some inputs to the system
and observe relevant outcomes.

The stronger white-box access level allows us to decompose the model
and trace an intermediate computation that is a proxy.
Such traceability is not afforded by the weaker access assumptions in
prior work. 
Thus, we explore a different point in the space by giving up on the
weaker access requirement to gain the ability to trace and repair
proxy use.

Tram\`{e}r et al.~\cite{fairtest} solve an important orthogonal
problem of efficiently identifying populations where associations may
appear. 
Since our definition is parametric in the choice of the population,
their technique could allow identifying relevant populations for
further analysis using our methods.

\paragraph{Repair}

Techniques for the repair of fairness violations are as varied as
fairness definitions.
Repair mechanisms that operate solely on the population dataset,
removing unfairness inherent in it, include variations that relabel
the class attribute~\cite{luong11knn}, modify entire instances while
maintaining the original schema~\cite{feldman15kdd}, and transform the
dataset into an alternate space of
features~\cite{zemel13icml,dwork12itcs}. 
Several approaches function instead on the training algorithm
employed, or rather introduce variations that ensure produced models
respect fairness constraints.
Repair in such approaches means replacing a standard algorithm with a
fairness-aware one, and requires access to the training data and the
learning pipeline (e.g.,~\cite{calders09dmwkshp, Calders2010,
  kamishima12euromlkdd}).
Adjustments to Naive Bayes~\cite{Calders2010} and trainers amiable to
regularization~\cite{kamishima12euromlkdd} are examples.

\section{Discussion}\label{sec:discussion}

Several design decisions and setting assumptions dictate where and how
our methodologies can be applied. 
We do not address the adversarial setting as our definition of proxy
use relies on strict program decomposition which can be subverted by
an intentional adversary. 
Further, our metrics of influence and association are based on a
distribution correlating program inputs with protected classes. 
What this distribution represents and how we to obtain it are both
points warranting discussion. 

\subsection{Beyond strict decomposition}\label{sub:discussion-strict}

Theorem~\ref{thm:sem-impossibility} shows that a definition satisfying
natural semantic properties is impossible.
This result motivates our syntactic definition, parameterized by a
programming language and a choice of program decomposition.
In our implementation, the choice of program decomposition is strict. 
It only considers single expressions in its decomposition. 
However, proxies may be distributed across different terms in the
program.
As discussed in Section~\ref{sec:detecting:model}, single expressions
decompositions can also deal with a restricted class of such
distributed proxies.
Our implementation does not identify situations where each of a large
number of syntactically different proxies have weak influence but
together combine to result in high influence.
A stronger notion of program decomposition that allows a collection of
multiple terms to be considered a proxy would identify such a case of
proxy use.

The choice of program decomposition also has consequences for the
tractability of the detection and repair algorithms.
The detection and repair algorithms summarized in this paper currently
enumerate through all possible subprograms in the worst case.
Depending on the flexibility of the language chosen and the
model\footnote{Though deep learning models can be expressed in the
  example language presented in this paper, doing so would result in
  prohibitively large programs.}
being expressed there could be an exponentially large number of
subprograms, and our enumeration would be intractable.

Important directions of future work are therefore organized along two
thrusts.
The first thrust is to develop flexible notions of program
decompositions that identify proxy uses for other kinds of machine
learning models, including deep learning models that will likely
require new kinds of abstraction techniques due to their size. 
The second thrust is to identify scalable algorithms for detecting and
repairing proxy use for these flexible notions of program
decomposition.

\subsection{Distributions and datasets}\label{sec:discussion-distribution}

As we noted starting in \sref{sec:notation}, our formalism is written
in terms of distributions whereas our practical implementation
operates on datasets.
Though it is best to think of a distribution as a sample of some
real-world population and thus an approximation of reality, we do not
address this assumption in our work.
Disparity between reality and the analyzed datasets introduces
concerns regarding the conclusions drawn from our methods.

If the analyzed dataset does not exhibit the associations establishing
proxies that do exist in the real world then we can no longer rely on
our methods to discover and repair real-world discriminatory
practices.
On the other hand, if a dataset introduces proxies that are not
present in the real world, subtle philosophical and legal questions
arise: does apparent discrimination still count as discrimination if
it does not apply to the real-world?
In the latter case of false positives, the ethical oracle in our
formalism can be used in place of philosophical or legal assessments.
The former possibility of a false negative, however, cannot be
resolved in our formalism as there are no regards for human
intervention without a proxy-use witness.

We believe that dataset disparity with reality is a problem orthogonal
to the issues we address in this work; our conclusions are only as
good as the accuracy of the datasets to which we apply our methods.

\subsection{Data requirements}\label{sec:discussion-data}

Our definitions and algorithms require access to datasets that contain
both necessary inputs to execute a model and the attributes indicating
protected class.
The latter may not be explicitly collected or inferred for use in the
models being analyzed.
Therefore, to discover unwanted proxy uses of protected classes, an
auditor might need to first infer the class from the collected data to
the best extent available to them.
If the protected class is sensitive or private information in the
specific context, it may seem ethically ambiguous to infer it in order
to (discover and) prevent its uses. 
However, this is consistent with the view that privacy is a function
of both information and the purpose for which that information is
being used \cite{TschantzDW12}\footnote{This principle is exemplified
  by law in various jurisdictions including the PIPEDA Act in Canada
  \cite{pipeda}, and the HIPAA Privacy Rule in the USA \cite{hipaa}.}.
In our case, the inference and use of the sensitive protected class by
an auditor has a different (and ethically justified) purpose than
potential inferences in model being audited.
Further, sensitive information has already been used by public and
private entities in pursuit of social good: affirmative action
requires the inference or explicit recording of minority membership,
search engines need to infer suicide tendency in order to show suicide
prevention information in their search results\cite{suicide}, health
conditions can potentially be detected early from search logs of
affected individuals \cite{paparrizos16screening}.
Supported by law and perception of public good, we think it justified
to expect system owners be cooperative in providing the necessary
information or aiding in the necessary inference for auditing.

\section{Conclusion}
\label{sect:conclusion}
We develop a theory of proxy discrimination in data-driven systems. 
Distinctively, our approach to use constrains not only the direct use
of protected class but also their proxies (i.e. 
strong predictors), unless allowed by exceptions justified by ethical
considerations.

We formalize proxy use and summarize a program analysis technique for
detecting it in a model. 
In contrast to prior work, our analysis is white-box. 
The additional level of access enables our detection algorithm to
provide a witness that localizes the use to a part of the algorithm.
Recognizing that not all instances of proxy use of a protected class
are inappropriate, our theory of proxy discrimination makes use of a
normative judgment oracle that makes this appropriateness
determination for a given witness. 
If the proxy use is deemed inappropriate, our repair algorithm uses
the witness to transform the model into one that does not exhibit
proxy use.
Using a corpus of social datasets, our evaluation shows that these
algorithms are able to detect proxy use instances that would be
difficult to find using existing techniques, and subsequently remove
them while maintaining acceptable classification performance.

\paragraph*{Acknowledgments}
We would like to thank Amit Datta, Sophia Kovaleva, and Michael C. 
Tschantz for their thoughtful discussions throughout the development
of this work.

 \begin{small}
   This work was developed with the support of NSF grants CNS-1704845,
   CNS-1064688 as well as by DARPA and the Air Force Research
   Laboratory under agreement number FA8750-15-2-0277. 
   The U.S. 
   Government is authorized to reproduce and distribute reprints for
   Governmental purposes not withstanding any copyright notation
   thereon. 
   The views, opinions, and/or findings expressed are those of the
   author(s) and should not be interpreted as representing the
   official views or policies of DARPA, the Air Force Research
   Laboratory, the National Science Foundation, or the U.S. 
   Government.
 \end{small}

\appendix

\bibliographystyle{plain}
\bibliography{%
  bibs/common.bib,%
  bibs/security.bib,%
  bibs/pl.bib,%
  bibs/other.bib,%
  bibs/causal-influence.bib,%
  bibs/blackbox.etc.bib,%
  bibs/legal.bib%
}

\begin{thebibliography}{10}

\bibitem{title-vii}
Title vii of the civil rights act of 1964, 1964.
\newblock (Accessed Aug 13, 2016).

\bibitem{ricci}
{Ricci et al. vs Destefano et al., US Supreme Court }, 2009.

\bibitem{ukequalityact2010}
Equality act of 2010, 2010.
\newblock (Accessed Jul 7, 2017).

\bibitem{adler2016auditing}
Philip Adler, Casey Falk, Sorelle~A Friedler, Gabriel Rybeck, Carlos
  Scheidegger, Brandon Smith, and Suresh Venkatasubramanian.
\newblock Auditing black-box models for indirect influence.
\newblock In {\em Data Mining (ICDM), 2016 IEEE 16th International Conference
  on}, pages 1--10. IEEE, 2016.

\bibitem{angwin16propublica}
Julia Angwin, Jeff Larson, Surya Mattu, and Lauren Kirchner.
\newblock Machine bias: There's software used across the country to predict
  future criminals. and it’s biased against blacks.
\newblock {\em ProPublica}, May 2016.

\bibitem{barocas2016big}
Solon Barocas and Andrew~D Selbst.
\newblock Big data’s disparate impact.
\newblock {\em California Law Review}, 104(3):671, 2016.

\bibitem{berk14forecasts}
Richard Berk and Justin Bleich.
\newblock Forecasts of violence to inform sentencing decisions.
\newblock {\em Journal of Quantitative Criminology}, 30(1):79--96, 2014.

\bibitem{berk16forecasting}
Richard~A. Berk, Susan~B. Sorenson, and Geoffrey Barnes.
\newblock Forecasting domestic violence: A machine learning approach to help
  inform arraignment decisions.
\newblock {\em Journal of Empirical Legal Studies}, 13(1):94--115, 2016.

\bibitem{Breiman01}
Leo Breiman.
\newblock Random forests.
\newblock {\em Mach. Learn.}, 45(1):5--32, October 2001.

\bibitem{calders09dmwkshp}
T.~Calders, F.~Kamiran, and M.~Pechenizkiy.
\newblock Building classifiers with independency constraints.
\newblock In {\em 2009 IEEE International Conference on Data Mining Workshops},
  pages 13--18, December 2009.

\bibitem{calders10naive}
Toon Calders and Sicco Verwer.
\newblock Three naive bayes approaches for discrimination-free classification.
\newblock {\em Data Mining and Knowledge Discovery}, 21(2):277--292, 2010.

\bibitem{Calders2010}
Toon Calders and Sicco Verwer.
\newblock Three naive bayes approaches for discrimination-free classification.
\newblock {\em Data Mining and Knowledge Discovery}, 21(2):277--292, 2010.

\bibitem{chickering00advertising}
David~Maxwell Chickering and David Heckerman.
\newblock A decision theoretic approach to targeted advertising.
\newblock In {\em Proceedings of the Sixteenth Conference on Uncertainty in
  Artificial Intelligence}, UAI'00, pages 82--88, San Francisco, CA, USA, 2000.
  Morgan Kaufmann Publishers Inc.

\bibitem{cortes95svm}
Corinna Cortes and Vladimir Vapnik.
\newblock Support-vector networks.
\newblock {\em Mach. Learn.}, 20(3):273--297, September 1995.

\bibitem{cover2012elements}
Thomas~M Cover and Joy~A Thomas.
\newblock {\em Elements of information theory}.
\newblock John Wiley \& Sons, 2012.

\bibitem{datta2015opacity}
A.~Datta, M.C. Tschantz, and A.~Datta.
\newblock Automated experiments on ad privacy settings: A tale of opacity,
  choice, and discrimination.
\newblock In {\em Proceedings on Privacy Enhancing Technologies (PoPETs 2015)},
  pages 92--112, 2015.

\bibitem{datta2015plsc}
Amit Datta, Anupam Datta, Deirdre Mulligan, and Michael Tschantz.
\newblock Discrimination in online personalization: A multidisciplinary
  inquiry.
\newblock {\em Privacy Law Scholars Conference}, 2015.
\newblock (Manuscript available upon request).

\bibitem{datta15pets}
Amit Datta, Michael~Carl Tschantz, and Anupam Datta.
\newblock Automated experiments on ad privacy settings: A tale of opacity,
  choice, and discrimination.
\newblock In {\em Proceedings on Privacy Enhancing Technologies (PoPETs)}. De
  Gruyter Open, 2015.

\bibitem{useprivacy}
Anupam Datta, Matthew Fredrikson, Gihyuk Ko, Piotr Mardziel, and Shayak Sen.
\newblock Use privacy in data-driven systems: Theory and experiments with
  machine learnt programs.
\newblock {\em CoRR}, abs/1705.07807, 2017.

\bibitem{qii}
Anupam Datta, Shayak Sen, and Yair Zick.
\newblock {Algorithmic Transparency via Quantitative Input Influence: Theory
  and Experiments with Learning Systems}.
\newblock In {\em Proceedings of IEEE Symposium on Security \& Privacy 2016},
  2016.

\bibitem{Dwork12}
C.~Dwork, M.~Hardt, T.~Pitassi, O.~Reingold, and R.~Zemel.
\newblock Fairness through awareness.
\newblock In {\em Proceedings of the 3rd Innovations in Theoretical Computer
  Science Conference (ITCS 2012)}, pages 214--226, 2012.

\bibitem{dwork12itcs}
Cynthia Dwork, Moritz Hardt, Toniann Pitassi, Omer Reingold, and Richard Zemel.
\newblock Fairness through awareness.
\newblock In {\em Proceedings of the 3rd Innovations in Theoretical Computer
  Science Conference}, ITCS '12, pages 214--226, New York, NY, USA, 2012. ACM.

\bibitem{dwork11fairness}
Cynthia Dwork, Moritz Hardt, Toniann Pitassi, Omer Reingold, and Richard~S.
  Zemel.
\newblock Fairness through awareness.
\newblock {\em Computing Research Repository (CoRR)}, 2011.

\bibitem{fed06book}
{Federal Reserve}.
\newblock {\em Consumer Compliance Handbook}, chapter Federal Fair Lending
  Regulations and Statutes: Overview.
\newblock Federal Reserve, 2016.

\bibitem{feldman15disparate}
Michael Feldman, Sorelle~A. Friedler, John Moeller, Carlos Scheidegger, and
  Suresh Venkatasubramanian.
\newblock Certifying and removing disparate impact.
\newblock In {\em Proceedings of the ACM SIGKDD International Conference on
  Knowledge Discovery and Data Mining (KDD)}, 2015.

\bibitem{feldman15kdd}
Michael Feldman, Sorelle~A. Friedler, John Moeller, Carlos Scheidegger, and
  Suresh Venkatasubramanian.
\newblock Certifying and removing disparate impact.
\newblock In {\em Proceedings of the 21th ACM SIGKDD International Conference
  on Knowledge Discovery and Data Mining}, KDD '15, pages 259--268, New York,
  NY, USA, 2015. ACM.

\bibitem{frees2014asbook}
Edward~W. Frees, Richard~A. Derrig, and Glenn Meyers.
\newblock {\em Predictive Modeling Applications in Actuarial Science}.
\newblock Cambridge University Press, 2014.

\bibitem{garg-jia-datta-ccs-2011}
Deepak Garg, Limin Jia, and Anupam Datta.
\newblock Policy auditing over incomplete logs: theory, implementation and
  applications.
\newblock In {\em Proceedings of the ACM Conference on Computer and
  Communications Security (CCS)}, 2011.

\bibitem{gepp16insurance}
Adrian Gepp, J.~Holton Wilson, Kuldeep Kumar, and Sukanto Bhattacharya.
\newblock A comparative analysis of decision trees vis-a-vis other
  computational data mining techniques in automotive insurance fraud detection.
\newblock {\em Journal of Data Science}, 10(3):537--561, 2012.

\bibitem{goguen1982security}
Joseph~A Goguen and Jos{\'e} Meseguer.
\newblock Security policies and security models.
\newblock In {\em Security and Privacy, 1982 IEEE Symposium on}, pages 11--11.
  IEEE, 1982.

\bibitem{hare03pclr}
Robert Hare.
\newblock {\em Manual For the Revised Psychopathy Checklist}.
\newblock Multi-Health Systems, 2003.

\bibitem{Hoeffding:1963}
Wassily Hoeffding.
\newblock Probability inequalities for sums of bounded random variables.
\newblock {\em Journal of the American Statistical Association},
  58(301):13--30, March 1963.

\bibitem{hunt2005}
D.~Bradford Hunt.
\newblock Redlining, 2005.
\newblock Encyclopaedia of Chicago.

\bibitem{ingold16bloomberg}
David Ingold and Spencer Soper.
\newblock {A}mazon doesn't consider the race of its customers. should it?
\newblock {\em Bloomberg}, April 2016.

\bibitem{kamishima12euromlkdd}
Toshihiro Kamishima, Shotaro Akaho, Hideki Asoh, and Jun Sakuma.
\newblock Fairness-aware classifier with prejudice remover regularizer.
\newblock In {\em Proceedings of the 2012 European Conference on Machine
  Learning and Knowledge Discovery in Databases - Volume Part II}, ECML
  PKDD'12, pages 35--50, Berlin, Heidelberg, 2012. Springer-Verlag.

\bibitem{kamishima11fairness}
Toshihiro Kamishima, Shotaro Akaho, and Jun Sakuma.
\newblock Fairness-aware learning through regularization approach.
\newblock In {\em Proceedings of the Workshop on Privacy Aspects of Data
  Mining}, 2011.

\bibitem{kennedy15scotus}
Anthony Kennedy.
\newblock Texas department of housing \& community affairs v. the inclusive
  communities project, inc.
\newblock Opinion of the United States Supreme Court, June 2015.

\bibitem{kilbertus2017avoiding}
Niki Kilbertus, Mateo Rojas-Carulla, Giambattista Parascandolo, Moritz Hardt,
  Dominik Janzing, and Bernhard Sch{\"o}lkopf.
\newblock Avoiding discrimination through causal reasoning.
\newblock {\em arXiv preprint arXiv:1706.02744}, 2017.

\bibitem{xray}
Mathias L{\'e}cuyer, Guillaume Ducoffe, Francis Lan, Andrei Papancea, Theofilos
  Petsios, Riley Spahn, Augustin Chaintreau, and Roxana Geambasu.
\newblock Xray: Enhancing the web's transparency with differential correlation.
\newblock In {\em Proceedings of the 23rd USENIX Conference on Security
  Symposium}, SEC'14, pages 49--64, Berkeley, CA, USA, 2014. USENIX
  Association.

\bibitem{sunlight}
Mathias Lecuyer, Riley Spahn, Yannis Spiliopolous, Augustin Chaintreau, Roxana
  Geambasu, and Daniel Hsu.
\newblock Sunlight: Fine-grained targeting detection at scale with statistical
  confidence.
\newblock In {\em Proceedings of the 22Nd ACM SIGSAC Conference on Computer and
  Communications Security}, CCS '15, pages 554--566, New York, NY, USA, 2015.
  ACM.

\bibitem{letham2015}
Benjamin Letham, Cynthia Rudin, Tyler~H. McCormick, and David Madigan.
\newblock Interpretable classifiers using rules and bayesian analysis: Building
  a better stroke prediction model.
\newblock {\em Ann. Appl. Stat.}, 9(3):1350--1371, 09 2015.

\bibitem{lichman2013uci}
M.~Lichman.
\newblock {UCI} machine learning repository, 2013.

\bibitem{lipton16privacy}
Richard~J. Lipton and Kenneth~W. Regan.
\newblock Making public information secret, 2016.
\newblock Accessed Aug 13, 2016.

\bibitem{luong11knn}
Binh~Thanh Luong, Salvatore Ruggieri, and Franco Turini.
\newblock k-nn as an implementation of situation testing for discrimination
  discovery and prevention.
\newblock In {\em Proceedings of the ACM SIGKDD International Conference on
  Knowledge Discovery and Data Mining (KDD)}, 2011.

\bibitem{marr15forbes}
Bernard Marr.
\newblock How big data is changing insurance forever.
\newblock {\em Forbes}, December 2015.

\bibitem{Meila2003}
Marina Meil{\u{a}}.
\newblock Comparing clusterings by the variation of information.
\newblock In {\em Proceedings of the Conference on Learning Theory and Kernel
  Machines (LTKM)}, 2003.

\bibitem{pipeda}
Office of~the Privacy Commissioner~of Canada.
\newblock Pipeda legislation and regulations, 2015.
\newblock Accessed May 15, 2017.

\bibitem{mcdonnell}
Supreme~Court of~United~States.
\newblock Mcdonnell douglas corp. v. green, 1973.

\bibitem{disparate-impact}
Supreme~Court of~United~States.
\newblock {E.G. Griggs v. Duke Power Co., 401 U.S. 424, 91 S. Ct. 849, 28 L.
  Ed. 2d 158}, 1977.

\bibitem{hipaa}
{Office for Civil Rights}.
\newblock Summary of the {HIPAA} privacy rule.
\newblock OCR Privacy Brief, U.S.\ Department of Health and Human Services,
  2003.

\bibitem{paparrizos16screening}
John Paparrizos, Ryen~W. White, and Eric Horvitz.
\newblock Screening for pancreatic adenocarcinoma using signals from web search
  logs: Feasibility study and results.
\newblock {\em Journal of Oncology Practice}, 12(8):737--744, 2016.
\newblock PMID: 27271506.

\bibitem{CausalityBook}
Judea Pearl.
\newblock {\em {Causality: Models, Reasoning and Inference}}.
\newblock Cambridge University Press, New York, NY, USA, 2nd edition, 2009.

\bibitem{rampton14reuters}
Roberta Rampton.
\newblock {W}hite {H}ouse looks at how `big data' can discriminate.
\newblock {\em Reuters}, April 2014.

\bibitem{rich2004performing}
Camille~Gear Rich.
\newblock Performing racial and ethnic identity: discrimination by proxy and
  the future of title vii.
\newblock {\em NYUL Rev.}, 79:1134, 2004.

\bibitem{suicide}
S.E. Smith.
\newblock How do search engines respond when you google ‘suicide’?, 2015.
\newblock Accessed May 15, 2017.

\bibitem{statt16verge}
Nick Statt.
\newblock {U}niversal {P}ictures made different {S}traight {O}utta {C}ompton
  trailers for different races: Delivered via {F}acebook.
\newblock {\em The Verge}, March 2016.

\bibitem{sweeney13cacm}
Latanya Sweeney.
\newblock Discrimination in online ad delivery.
\newblock {\em Commun. ACM}, 56(5):44--54, May 2013.

\bibitem{fairtest}
Florian Tram{\`{e}}r, Vaggelis Atlidakis, Roxana Geambasu, Daniel~J. Hsu,
  Jean{-}Pierre Hubaux, Mathias Humbert, Ari Juels, and Huang Lin.
\newblock Discovering unwarranted associations in data-driven applications with
  the fairtest testing toolkit.
\newblock {\em CoRR}, abs/1510.02377, 2015.

\bibitem{tschantz12sp}
Michael~Carl Tschantz, Anupam Datta, and Jeannette~M. Wing.
\newblock Formalizing and enforcing purpose restrictions in privacy policies.
\newblock In {\em Proceedings of the 2012 IEEE Symposium on Security and
  Privacy}, pages 176--190, Washington, DC, USA, 2012.

\bibitem{TschantzDW12}
Michael~Carl Tschantz, Anupam Datta, and Jeannette~M. Wing.
\newblock Formalizing and enforcing purpose restrictions in privacy policies.
\newblock In {\em {IEEE} Symposium on Security and Privacy, {SP} 2012, 21-23
  May 2012, San Francisco, California, {USA}}, pages 176--190, 2012.

\bibitem{statistics}
Graham Upton and Ian Cook.
\newblock {\em Dictionary of Statistics}.
\newblock Oxford University Press, 2008.

\bibitem{zemel13fair}
Rich Zemel, Yu~Wu, Kevin Swersky, Toni Pitassi, and Cynthia Dwork.
\newblock Learning fair representations.
\newblock In {\em Proceedings of the Internetional Conference on Machine
  Learning}, 2013.

\bibitem{zemel13icml}
Richard~S. Zemel, Yu~Wu, Kevin Swersky, Toniann Pitassi, and Cynthia Dwork.
\newblock Learning fair representations.
\newblock In {\em Proceedings of the 30th International Conference on Machine
  Learning}, volume~28 of {\em {JMLR} Workshop and Conference Proceedings},
  pages 325--333. JMLR.org, 2013.

\end{thebibliography}

\end{document}